\documentclass[twocolumn,apj,iop,numberedappendix,a4paper]{openjournal}
\usepackage{amsmath}
\usepackage{amssymb}
\usepackage{verbatim}
\usepackage{graphicx}
\usepackage{booktabs}
\usepackage{array}
\usepackage[dvipsnames]{xcolor}
\usepackage{xspace}
\usepackage{ulem}
\usepackage{lipsum}
\usepackage{xfrac}
\usepackage[colorlinks=true, allcolors=blue]{hyperref}
\usepackage{natbib}
\usepackage{ulem}

\newcommand{\class}{\texttt{class}}
\newcommand{\mgclass}{\texttt{mgclass}}
\newcommand{\mochiclass}{\texttt{mochi\_class}}
\newcommand{\hiclass}{\texttt{hi\_class}}
\newcommand{\camb}{\texttt{CAMB}}
\newcommand{\eftcamb}{\texttt{EFTCAMB}}
\newcommand{\mgcamb}{\texttt{MGCAMB}}
\newcommand{\leftcap}{\textit{Left:}}
\newcommand{\rightcap}{\textit{Right:}}
\newcommand{\lcdm}{$\Lambda$CDM}
\newcommand{\proptoomega}{\texttt{propto\_omega}}
\newcommand{\fr}{$f(R)$}
\newcommand{\Dkin}{D_{\rm kin}}
\newcommand{\DeltaMpl}{\Delta M_\ast^2}
\newcommand{\Mpl}{M_\ast^2}
\newcommand{\alphaT}{\alpha_{\rm T}}
\newcommand{\alphaM}{\alpha_{\rm M}}
\newcommand{\alphaB}{\alpha_{\rm B}}
\newcommand{\alphaK}{\alpha_{\rm K}}
\newcommand{\cs}{c_{\rm s}^2}
\newcommand{\csnum}{c_{\rm sN}^2}
\newcommand{\rhophi}{\rho_\phi}
\newcommand{\rhom}{\rho_{\rm m}}

\newcommand{\presm}{p_{\rm m}}
\newcommand{\presphi}{p_{\phi}}
\newcommand{\normrhophi}{\tilde{\rho}_\phi}
\newcommand{\alphaBic}{\alpha_{{\rm B}0}}
\newcommand{\hMpc}{h/{\rm Mpc}}
\newcommand{\Om}{\Omega_{\rm m}}

\newcommand{\diff}{{\rm d}}

\DeclareMathOperator{\sign}{sgn}

\begin{document}

\title{\mochiclass{}: Modelling Optimisation to Compute Horndeski in \class{}}

\author{M. Cataneo$^{1,2,*}$}
\author{E. Bellini$^{3,4,5}$}
\email{$^*$mcataneo@uni-bonn.de}
\affiliation{$^1$Ruhr University Bochum, Faculty of Physics and Astronomy, Astronomical Institute (AIRUB), German Centre for Cosmological Lensing, 44780 Bochum, Germany}
\affiliation{$^2$Argelander-Institut für Astronomie (AIfA), Universität Bonn, Auf dem Hügel 71, 53121 Bonn, Germany}
\affiliation{$^3$SISSA, International School for Advanced Studies, Via Bonomea 265, 34136 Trieste, Italy}
\affiliation{$^4$IFPU, Institute for Fundamental Physics of the Universe, via Beirut 2, 34151 Trieste, Italy}
\affiliation{$^5$INFN, National Institute for Nuclear Physics, Via Valerio 2, I-34127 Trieste, Italy}


\begin{abstract}
    We introduce \mochiclass{}, an extension of the Einstein-Boltzmann solver \hiclass{}, designed to unlock the full phenomenological potential of Horndeski gravity. This extension allows for general input functions of time without the need for hard-coded parametrisations or covariant Lagrangians. By replacing the traditional \(\alpha\)-parametrisation with a set of stable basis functions, \mochiclass{} ensures that the resulting effective theories are inherently free from gradient and ghost instabilities. Additionally, \mochiclass{} features a quasi-static approximation implemented at the level of modified metric potentials, enhancing prediction accuracy, especially for models transitioning between a super- and sub-Compton regime. \mochiclass{} can robustly handle a wide range of models without fine-tuning, and introduces a new approximation scheme that activates modifications to the standard cosmology deep in the matter-dominated era. Furthermore, it incorporates viability conditions on the equation of motion for the scalar field fluctuations, aiding in the identification of numerical instabilities. Through comprehensive validation against other Einstein-Boltzmann solvers, \mochiclass{} demonstrates excellent performance and accuracy, broadening the scope of \hiclass{} by facilitating the study of specific modified gravity models and enabling exploration of previously inaccessible regions of the Horndeski landscape. The code is publicly available at \href{https://github.com/mcataneo/mochi_class_public}{\texttt{https://github.com/mcataneo/mochi\_class\_public}}.
\end{abstract}

\maketitle

\section{Introduction}
\label{sec:intro}

The current cosmological paradigm ($\Lambda$CDM) is capable of describing the dynamics of the universe, the Cosmic Microwave Background (CMB) and the Large-Scale Structure (LSS) of the universe with only six parameters~\citep{Aghanim2018}. On top of standard model particles, it assumes two exotic components whose nature is unknown: (i) a cosmological constant ($\Lambda$) and some form of Cold Dark Matter (CDM). The gravitational interactions are regulated by General Relativity (GR). This modelling explains the observed accelerated expansion of the universe~\citep{Riess1998,Perlmutter1998}, and it provides a good fit to the data~\citep[e.g.,][]{Heymans2020,Alam2020,DES2021,Miyatake2023}.

However, the standard cosmological model is unsatisfactory as it suffers from at least three drawbacks: (i) the value of the observed cosmological constant is too small to be explained by fundamental physics~\citep{Martin2012}, (ii) GR is tested very precisely on local scales~\citep{Will2014} but it is extrapolated over 15 orders of magnitude in length scale to be applied to cosmological data~\citep{Peebles2004}, and (iii) comparing different datasets produces tensions for the measured parameters~\citep{Valentino2021,Abdalla2022}. In particular, the most significant is the Hubble tension, arising when combining ``early'' and ``late'' times measurements of the expansion rate of the universe \citep{Verde2019}.

To alleviate these problems several possible extensions have been proposed, either promoting $\Lambda$ to a dynamical field (Dark Energy, DE)~\citep[see, e.g.,][for a review]{Copeland2006} or modifying the laws of gravity at cosmological scales (MG)~\citep[see, e.g.,][for reviews]{Clifton2011,Joyce2014,Joyce2016}. A unifying framework is realised by the Horndeski theory, that encompasses many of the scalar-tensor theories proposed in the literature, e.g.~Quintessence, $f(R)$ gravity, Brans-Dicke, Galileons.
Horndeski gravity provides one of the most general scalar-tensor theories with at most second-order derivatives in the equations of motion on any background \citep{Horndeski1974,Deffayet2009a,Kobayashi2011}. It gained significant popularity due to its rich phenomenology while remaining analytically tractable.

To study the Horndeski landscape, there are two potential strategies~\citep[e.g.,][]{Ezquiaga2018}. It is possible to: (i) select a specific model based on its covariant Lagrangian formulation and analyse the effects of the additional degree of freedom across different regimes, from cosmological scales to compact objects and black holes; or (ii) adopt an effective-theory approach (EFT), focusing solely on the energy scales relevant to cosmology.

While strategy (i) allows for a detailed investigation of particular scalar-tensor theories, it demands significant resources and time investment. In the absence of a compelling alternative to GR, strategy (ii) provides a more general and efficient method to detect or constrain deviations from standard gravity~\citep[e.g.,][]{Ezquiaga2017,Creminelli2017,Baker2017,Sakstein2017}.

Given the large classes of models that the Horndeski Lagrangian encapsulates, there have been countless attempts to get cosmological constraints on specific parametrisations. Following each one of the two distinct approaches described above, in the literature it is possible to find both constraints based on sub-classes of Horndeski~\citep[see, e.g.,][]{Joudaki2022, Zumalacarregui2020, Ye2024}, and on different parametrisations of the EFT functions~\citep[see, e.g.,][]{Bellini2015a, Kreisch2017, Noller2018, Mancini2019, Seraille2024, Castello2024, Chudaykin2024}. A promising but less explored direction is to get model independent constraints. The idea is to try to ``reconstruct'' gravity without imposing an a priori time dependence for the EFT functions \citep{Zhao2009, Zhao2010, Hojjati2012, Raveri2019, Pogosian2021, Raveri2021, Park2021}.

The public version of \hiclass{} \citep{Zumalacarregui2016, Bellini2019}, an extension of the Einstein-Botzmann solver \class{} \citep{Blas2011, Lesgourgues2011c}, implements both approaches, allowing users to easily implement new models belonging to Horndeski scalar-tensor theories or the EFT framework with the basis proposed in \cite{Bellini2014}. In a second release of the code, \cite{Bellini2019} included the quasi-static approximation scheme for improved computational efficiency and to extend the code's applicability to models where following the full dynamics of the scalar field can be extremely challenging.

Even with these significant advances, further steps are required to reliably extend \hiclass{} across a broader range of the Horndeski landscape. First, we need to simplify the implementation of specific models and parametrisations, which currently requires modifications to the source code. Additionally, efficient selection of stable models is crucial to avoid extensive searches for parameter combinations free from ghost and gradient instabilities. The accuracy of the QSA implementation in \hiclass{} can be improved. Furthermore, addressing numerical noise from limited precision in the calculation of the scalar field speed of sound is essential to prevent incorrectly labeling healthy models as unstable.

To address these issues, in this paper we introduce \mochiclass{}, a numerical tool to facilitate the study and statistical analysis of Horndeski models expressed in the effective-theory language. To improve flexibility, \mochiclass{} externalises the models definition, allowing to pass to the code precomputed arrays for the time evolution of the background and the EFT functions. In addition, it implements a new stable parametrisation, to satisfy the stability conditions by construction. Furthemore, we implement a new integrated approximation scheme to switch on gravity modifications only after a certain (user specified) redshift. This fixes the early time evolution to $\Lambda$CDM, avoiding numerical instabilities when deviations from standard gravity are irrelevant. Finally, \mochiclass{} implements an additional QSA approximation implemented at the level of modified metric potentials in Newtonian gauge. This offers a cross-check with the standard \hiclass{} implementation and improves the agreement with the exact solution.

Throughout the paper we assume the speed of light $c=1$, use the \class{} normalisation for the components of the stress-energy tensor (e.g., 3$\rho_{\texttt{class}} = 8\pi G\rho_{\rm standard}$), and adopt the following values for the cosmological parameters: for the fractional energy-density of the cold dark matter and baryons, we set $\Omega_{\rm c}h^2 = 0.120108$ and $\Omega_{\rm b}h^2 = 0.022383$; for the Hubble constant, $h = 0.6781$; for the amplitude and slope of the primordial power spectrum, $10^9 A_{\rm s} = 2.10055$ and $n_{\rm s} = 0.96605$; for the optical depth at reionisation, $\tau = 0.054308$. In what follows we also assume that the background is described by a flat Friedmann-Lema\^itre-Robertson-Walker (FLRW) metric with signature ($-$+++).

The paper is organised as follow. In Section \ref{sec:horndeski} we introduce the Horndeski models and their EFT description. In Section \ref{sec:code_description} we describe the code structure and implementation, and in Section \ref{sec:code_validation} we validate it against other codes. In Section \ref{sec:new_models} we present a new parametrisation, and leave Section \ref{sec:summary} for the summary and outlook.

\section{Horndeski's theory and parametrisations}\label{sec:horndeski}

In our analysis we focus on sub-classes of the Horndeski theory, one of the most general scalar-tensor theories with at most second-order equations of motion on any background \citep{Horndeski1974,Deffayet2009a,Kobayashi2011}. Its action can be written as
\begin{subequations}
\begin{equation} \label{eq:full_horndeski}
  S_{\rm H}[g_{\mu\nu},\phi]=\int\mathrm{d}^{4}x\,\sqrt{-g}\left[\sum_{i=2}^{5}\frac{1}{8\pi
                            G_{\mathrm{N}}}{\cal L}_{i}[g_{\mu\nu},\phi] + \mathcal{L}_m\right]\,,
\end{equation}
\begin{align}
  &{\mathcal{L}}_{2} = G_{2}(\phi,\,X)\,, \\
  &{\mathcal{L}}_{3} = -G_{3}(\phi,\,X)\Box\phi\,, \\
  &{\mathcal{L}}_{4} = G_{4}(\phi,\,X)R+G_{4X}(\phi,\,X)\left[\left(\Box\phi\right)^{2}-\phi_{;\mu\nu}\phi^{;\mu\nu}\right]\,, \\
  &{\mathcal{L}}_{5} = G_{5}(\phi,\,X)G_{\mu\nu}\phi^{;\mu\nu} \nonumber \\
                        & \quad\quad -{\frac{1}{6}}G_{5X}(\phi,\,X)\left[ (\Box\phi)^{3} + 2\phi_{;\mu}{}^{\nu}\phi_{;\nu}{}^{\alpha}\phi_{;\alpha}{}^{\mu} \right.  \nonumber \\ 
                        & \quad\quad \left.{} - 3\phi_{;\mu\nu}\phi^{;\mu\nu}\Box\phi \right].
\end{align}
\end{subequations}
Here, $g_{\mu\nu}$ is the metric tensor, $\mathcal{L}_m$ is the matter Lagrangian and $G_{2,\,3,\,4,\,5}$ are arbitrary functions of the scalar field $\phi$ and its canonical kinetic term $X=-\phi^{;\mu}\phi_{;\mu}/2$.

In a cosmological setup, there are two complementary approaches that are commonly used. Each one has its own advantages and motivations. For completeness we quickly review both, as it will be useful for the rest of the paper.

A first approach consists in directly specifying the field dependence of the $G_{2,\,3,\,4,\,5}$ functions. This is the case of e.g.~quintessence or $f(R)$ gravity. Assuming a FLRW background one can get the evolution of the matter and the scalar field at the background level. Then, it is possible to get the evolution of the perturbations on top of the background obtained. The key advantage of this approach is \textit{self-consistency}. For a given covariant Lagrangian formulation one can derive the necessary equations in every regime (background vs perturbations, weak vs strong gravity) and apply the result of one regime to the others. This is extremely important, since the laws of gravity and DE should be universal. However, it is rather cumbersome to jump from one model to another, because the background evolution is non trivial and the effort to setup the machinery needed to solve the physical system under investigation has to be repeated for each model. Then, the main limitation of this approach is that without a favorite model in mind it is difficult to scan a significant part of the parameter space of gravity.

Alternatively it is possible to use an Effective Field Theory (EFT) approach. This framework compresses the information content of the full Equation~\eqref{eq:full_horndeski} into few functions of time. Assuming a FLRW background and up to linear order in perturbation theory a common choice is to describe the linear evolution of cosmological perturbations in Horndeski with \citep{Bellini2014}
\begin{align}\label{eq:alphas}
    & \{w_\phi,\,\alphaK,\,\alphaB,\,\alphaT,\,M_\ast^2\}\,.
\end{align}
Here, $w_\phi=p_\phi/\rho_\phi$ is the equation of state of the scalar field (or we can use any other function describing the background evolution of DE), $\alpha_{\rm K}$ has been dubbed \textit{kineticity}, $\alpha_{\rm B}$ is called \textit{braiding}, $\alpha_{\rm T}$ is the \textit{tensor-speed excess} and $M_*^2$ an effective \textit{Planck mass}. $M_*^2$ can also be replaced with its run-rate
\begin{equation}\label{eq:run}
    \alphaM \equiv \frac{\diff \ln M_\ast^2}{\diff \ln a} \, ,
\end{equation}
together with the initial condition $M_{\ast,{\rm ini}}$. Here and throughout, $a$ represents the scale factor of the FLRW metric. These functions describe all the possible dynamics of the Horndeski Lagrangian and they are independent of each other. $w_\phi$ is the only function affecting the background evolution (and the perturbations through a different expansion history), while the other functions in Equation~\eqref{eq:alphas} modify only the evolution of the perturbations. As a consequence, a commonly used strategy is to fix the background evolution to the one predicted by some fiducial model (typically $\Lambda$CDM) and focus on the phenomenology of the perturbations.
In $\Lambda$CDM the EFT functions in Equation~\ref{eq:alphas} take the values
\begin{align}\label{eq:lcdm_limit}
    & \{w_\phi=-1,\,\alphaK=0,\,\alphaB=0,\,\alphaT=0,\,M_*^2=1\}\,.
\end{align}
One of the key advantages of the EFT approach is that it is very powerful to test deviations from the standard cosmology. Indeed, the usual (but not unique) approach is to choose a time dependence of the EFT functions such that $\Lambda$CDM is recovered at early times. The EFT framework can be seen as a bridge between the underlying Horndeski theory and observations (see Figure 1 of \citealp{Bellini2019}), thus facilitating the inference of cosmological constraints. Then, for a given time evolution of the EFT functions it is possible to get a particular Horndeski covariant model following the strategy introduced in \citep{Kennedy2017, Kennedy2018}.

In particular, we choose to work within a sub-class of Horndeski models, specifically those that have $\alphaT = 0$. The nearly-simultaneous detection of the Gravitational Wave (GW) signal (GW170817) and the corresponding gamma-ray burst (GRB 170817A) emitted from the merger of a binary neutron star system \citep{LIGO2017} suggests that the speed of GW ($c_T^2\equiv 1+ \alphaT$) has to be equal to the speed of light for any cosmological purpose \citep{Baker2017, Creminelli2017, Ezquiaga2017, Sakstein2017}. While we make this assumption for simplicity, it is important to stress that there are a number of caveats that may invalidate it. In particular the speed of tensors can be frequency dependent, while the LIGO/VIRGO detection only probes $>\mathcal{O}(1){\rm Hz}$ frequencies. Then, it is possible to accommodate all observations by assuming a time varying cosmological $\alphaT$ and $\alphaT=0$ on LIGO/VIRGO frequencies. In addition, one should discuss about the validity of the EFT framework itself at the LIGO scales \citep{deRham2018}. However, given that this has to be intended as a first exploration of model optimization with \mochiclass{}, we prefer to keep the parameter space under investigation as little as possible.

Translating this requirement to the Horndeski functions implies $G_4=G_4\left(\phi\right)$ and $G_5=0$. Then Equation~(\ref{eq:full_horndeski}) becomes
\begin{equation} \label{eq:scalar_horndeski}
  {\mathcal{L}}_{\mathrm{SH}} = G_{2}(\phi,X)+G_{4}(\phi)R-G_{3}(\phi,X)\Box\phi\,,
\end{equation}
where SH stands for {\it Scalar Horndeski}.

As explained above, the background of this class of models can be fully specified by one function of time. It can be the scalar field energy-density, the scalar field equation of state, or directly the Hubble rate. This property is manifest by construction when working within the EFT framework. If working with the Horndeski $G_i$ functions one can project them into the background function chosen~\citep[see, e.g., Equations (A.6) and (A.7) in][]{Zumalacarregui2016}.

The evolution of the perturbations depends on both the chosen expansion history and on the other EFT functions in a non-trivial way. An in depth discussion that adopts the same notation used here can be found in, e.g.,~\cite{Bellini2014}. Here it is important for the rest of the discussion to report the definition of the de-mixed kinetic term for the scalar field perturbation,
\begin{equation}\label{eq:Dkin}
    \Dkin \equiv \alphaK + \frac{3}{2}\alphaB^2 \, ,
\end{equation}
and of its effective sound speed
\begin{align}\label{eq:cs2}
    c_{\mathrm{s}}^{2} = \frac{1}{\Dkin} & \left[(2-\alpha_{\mathrm{B}}) \left(-{\frac{H^{\prime}}{a H^{2}}}+{\frac{1}{2}}\alpha_{\mathrm{B}}+\alpha_{\mathrm{M}}\right) \right. \nonumber \\ 
    & \qquad\qquad\qquad \left. -{\frac{3\left(\rho_{\mathrm{m}}+p_{\mathrm{m}}\right)}{H^{2}M_{\ast}^{2}}}+{\frac{\alpha_{\mathrm{B}}^{\prime}}{a H}}\right] \, ,
\end{align}
where $H$ is the Hubble parameter, $\rhom$ and $\presm$ are the energy-density and pressure, respectively, of the matter species (excluding the scalar field), and primes denote conformal time derivatives. A definition of the $\alpha_i$'s as a function of the Horndeski $G_i$'s can be found in, e.g.,~\cite{Bellini2014}.

The models we are going to focus on are:
\\\\
\emph{Cubic Covariant Galileon.}
The covariant Galileon \citep{Deffayet2009b} model corresponds to the sub-class of Horndeski theories, Equation~(\ref{eq:full_horndeski}), that possesses the Galilean symmetry in a flat spacetime, i.e.~$\partial_\mu\phi\rightarrow\partial_\mu\phi+b_\mu$, with $b_\mu$ a constant 4-vector~\citep{Nicolis2008}. While the most general version of the covariant Galileon does not respect the condition $\alphaT=0$, here we focus on the cubic Galileon case, for which the $G_i$ functions read
\begin{equation}\label{eq:cubic_gal_lagrangian}
    G_{2} = c_2 X\,, \qquad G_{3}=2\frac{c_3}{\Lambda_3^3}X\,, \qquad G_{4}=\frac{M_{\rm Pl}^2}{2}\,.
\end{equation}
Here, as usual, we have set $\Lambda_{3}^{3}\equiv H_0^2 M_{\rm Pl}$. On top of the initial conditions (IC) for the scalar field $\phi$ and its time derivative $\dot\phi \equiv {\rm d}\phi/{\rm d}t$, $c_2$ and $c_3$ are extra parameters. In particular, $c_2$ can be normalized to $\pm 1$ through a convenient redefinition of the scalar field. The sign of $c_2$ is crucial for the expansion history of the universe. If $c_2=+1$ one needs a cosmological constant to drive the accelerated expansion of the universe. On the contrary, if $c_2=-1$ the model self-accelerates. Since the model is shift symmetric, the IC chosen for the scalar field itself is irrelevant, while the IC for its derivative can be fixed by following the attractor solution $H\dot\phi={\rm const.}$~\citep{Felice2010a}. Additionally, $c_3$ is fixed by requiring that the sum of the fractional energy-densities today is equal to one in a flat universe. Finally, we choose the self-accelerating branch ($c_2 = -1$), which gives
\begin{align}
    & \alphaK=3\alphaB=\frac{6\Lambda_{3}^{6}\Omega_{\phi,0}}{M_{\rm Pl}^2H^4}\,,\qquad M_*^2=1\,,\qquad \alphaM=0\,,
\end{align}
where $\Omega_{\phi,0}$ is the fractional energy-density of the scalar field today, and the Hubble rate is the only time-dependent quantity.

\emph{Kinetic Gravity Braiding.}
The Kinetic Gravity Braiding (KGB) model in its original formulation \citep{Deffayet2010} is a cubic Horndeski theory. Here we focus on a model (nKGB) defined as
\begin{equation}\label{eq:nkgb_lagrangian}
    G_{2}=-X\,,\qquad G_{3}=g^{(2n-1)/2} \frac{X^n}{\Lambda^{4n-1}}\,,\qquad G_{4}=\frac{M_{\rm Pl}^2}{2}\,,
\end{equation}
where $\Lambda^{(4n-1)}\equiv M_{\rm Pl}^{(2n-1)} H_0^{2n}$. The sign of the standard kinetic term has been chosen negative, to have a self-accelerating universe. $g$ and $n$ are the two free parameters of the theory, and the exponent of $g$ has been chosen such that $g\Omega_\phi(a=1)\sim \mathcal{O}(1)$ for every choice of $n$. It is easy to see that Equation~\eqref{eq:nkgb_lagrangian} is a generalisation of Equation~\eqref{eq:cubic_gal_lagrangian}, obtained by fixing $n=1$ and $g=4c_3^2$. In nKGB the only non-zero $\alpha_i$ functions are
\begin{align}
    &\alphaK = \frac{2 X}{M_{\rm Pl}^2 H^2}\left[- 1 + 6 n^2 g^{n-\frac{1}{2}} \frac{H \dot{\phi} X^{n-1}}{\Lambda^{4 n - 1}}\right] \, ,\\
    &\alphaB = 2 n \frac{g^{n-\frac{1}{2}} \dot{\phi} X^n}{\Lambda^{4 n - 1} M_{\rm Pl}^2 H} \, ,
\end{align}
and here overdots represent derivatives with respect to cosmic time. In our analysis we fix $n=5$, while $g$ is obtained from requiring the sum of the fractional energy densities of all species to be one today.

\emph{\fr{} gravity.} This scalar-tensor theory extends the Einstein-Hilbert action by adding a non-linear function of the Ricci scalar, \( f(R) \). After defining \(\phi \equiv 1 + f_R\), with \(f_R \equiv \mathrm{d}f/\mathrm{d}R\), it can be reformulated in terms of the Horndeski \(G_i\) functions as follows~\citep[see, e.g.,][]{Felice2010}:
\begin{equation}\label{eq:fr_lagrangian}
    G_{2} = \frac{f - R f_R}{2} \, , \qquad G_{3} = 0 \, , \qquad G_{4} = \frac{1 + f_R}{2} \, .
\end{equation}
From the definition of the \(\alpha_i\) functions~\citep{Bellini2014}, we can immediately see that:
\begin{equation}\label{eq:fr_alphas}
    \Mpl = 1 + f_R \, , \quad \alphaM = \frac{\dot{f}_R}{1 + f_R} \, , \quad \alphaB = -\alphaM \, , \quad \alphaK = 0 \, .
\end{equation}
Note that here and in the remainder of this paper overdots denote derivatives with respect to the natural logarithm of the scale factor.

The evolution of the scalar field and the background expansion are related by the modified Friedmann equation~\citep[see, e.g.,][]{Hu2007}
\begin{equation}\label{eq:fr_friedmann}
    H^{2}-f_{R}(H \dot H + H^{2})+\frac{1}{6}f+H^{2}f_{R R} \dot R= \rhom \, ,
\end{equation}
coupled with the equation for the Ricci scalar
\begin{equation}\label{eq:ricci}
    R = 12 H^2 + 6 H \dot H \, .
\end{equation}
In Equation~\eqref{eq:fr_friedmann}, $f_{RR} = {\rm d}f_R/{\rm d}R$, and $\rhom$ includes all matter species, that is, radiation, non-relativistic matter, neutrinos etc. 

Thus, we can either specify the expansion history and solve for $f(R)$, or provide an explicit functional form for the extension to the Einstein-Hilbert action and derive the associated background evolution. In the first case, we adopt the so-called {\it designer} approach~\citep{Song2006,Pogosian2007}, and in this work, we fix the background to that of a \lcdm{} cosmology. For this family of models, deviations from standard gravity are generally parametrised in terms of the dimensionless background quantity
\begin{equation}
    B = \frac{\dot f_R}{1 + f_R}\frac{H}{\dot H} \, ,
\end{equation}
and we will consider a model with $B_0 \equiv B(z = 0) = 0.01$.

Alternatively, we can adopt an explicit Lagrangian formulation, with the high-curvature limit of the \cite{Hu2007} form
\begin{equation}\label{eq:hu_sawicki}
    \frac{f(R)}{m^2} = -\frac{c_{1}}{c_{2}} + \frac{c_{1}}{c_{2}^{2}} \left({\frac{m^{2}}{R}}\right)^{n} \, ,
\end{equation}
being one of the most extensively studied models. Here $c_1$, $c_2$ and $n$ are free dimensionless parameters, and the mass scale $m^2 \equiv \rho_{\rm c} + \rho_{\rm b}$, where $\rho_{\rm c}$ and $\rho_{\rm b}$ are the background energy-density of cold dark matter and baryons, respectively. We can eliminate one parameter by requiring convergence to a cosmological constant in the early universe (i.e $R \gg m^2$), which gives $c_1/c_2 = 6\tilde\Omega_\Lambda/\Omega_{\rm m}$, with $\tilde\Omega_\Lambda = \lambda \Omega_\Lambda$, and $\lambda$ is a real positive number that accounts for deviations from the late-time value $\Omega_\Lambda = 1 - \Omega_{\rm m} - \Omega_{\rm r}$. We can also express the combination $c_1/c_2^2$ in terms of the initial scalar field value, $f_{R,{\rm ini}}$, allowing Equation~\eqref{eq:hu_sawicki} to be rewritten as
\begin{equation}
    \frac{f(R)}{m^2} = -6\frac{\tilde\Omega_\Lambda}{\Omega_{\rm m}} - \frac{f_{R,{\rm ini}}}{n}\left( \frac{R_{\rm ini}}{m^2} \right) \left( \frac{R_{\rm ini}}{R} \right)^n \, ,
\end{equation}
where $R_{\rm ini} \equiv R(a_{\rm ini})$.

We consider models with $n=1$ and $n=4$, and adjust $\lambda$ and $f_{R,{\rm ini}}$ such that at $z=0$ we have $\rho_{\phi} = \Omega_\Lambda H_0^2$ and $|f_{R0}| = 10^{-4}$~\footnote{While this $f_{R0}$ value is still consistent with the cosmological constraints obtained from linear scales~\citep{Hu2016}, it also ensures the perturbations can be computed in a few minutes when following the full dynamics of the scalar field fluctuations.}. We found $\lambda \sim \mathcal{O}(1 \pm f_{R0})$ (see, e.g., Figure~\ref{fig:rho_phi_reconstruction} in Section~\ref{sec:reconstruction}). To integrate the coupled system of Equations~\eqref{eq:fr_friedmann} and~\eqref{eq:ricci} we follow the strategy discussed in~\cite{Hu2016}. Finally, note that by combining Equations~\eqref{eq:fr_alphas} and \eqref{eq:fr_friedmann} into Equation~\eqref{eq:cs2} one finds the well-known result $\cs = 1$ valid for any $f(R)$ gravity model~\citep[see, e.g.,][]{Lombriser2018}.

\emph{Fixed-form parametrisation.}
Using the EFT approach we use a simple, yet effective, parametrisation. The background is fixed to $\Lambda$CDM, i.e.~$w=-1$, and the alpha functions defined as
\begin{equation}\label{eq:propto_omega}
    \alpha_i = \hat\alpha_i \Omega_\Lambda(a)\,.
\end{equation}
Here $\Omega_\Lambda(a)$ is the fractional energy density of the cosmological constant, and the $\hat\alpha_i$'s are free parameters that fix the amplitude of the EFT functions. The idea is to have a $\Lambda$CDM universe before the onset of DE, and allow for modifications at late times. Equation~(\ref{eq:propto_omega}) does not pretend to emulate realistic Horndeski models (with the exception of simple Quintessence), but rather to provide a minimal extension capable of capturing the phenomenology of all the alpha functions.

Given that the background evolution is fixed to be $\Lambda$CDM, on top of the standard cosmological parameters the only extra ones are
\begin{align}
    & \{\hat\alpha_{\rm K},\,\hat\alpha_{\rm B},\,\hat\alpha_{\rm M}\}\,, 
\end{align}
to which we assign $\hat\alpha_{\rm K}=1$, $\hat\alpha_{\rm B}=0.2$ and $\hat\alpha_{\rm M}=0.1$ for our test cosmology.

\subsection{Stable parametrisation}
\begin{figure*}[ht]
    \centering
    \begin{minipage}{.49\textwidth}
        \centering
        \includegraphics[width=\linewidth]{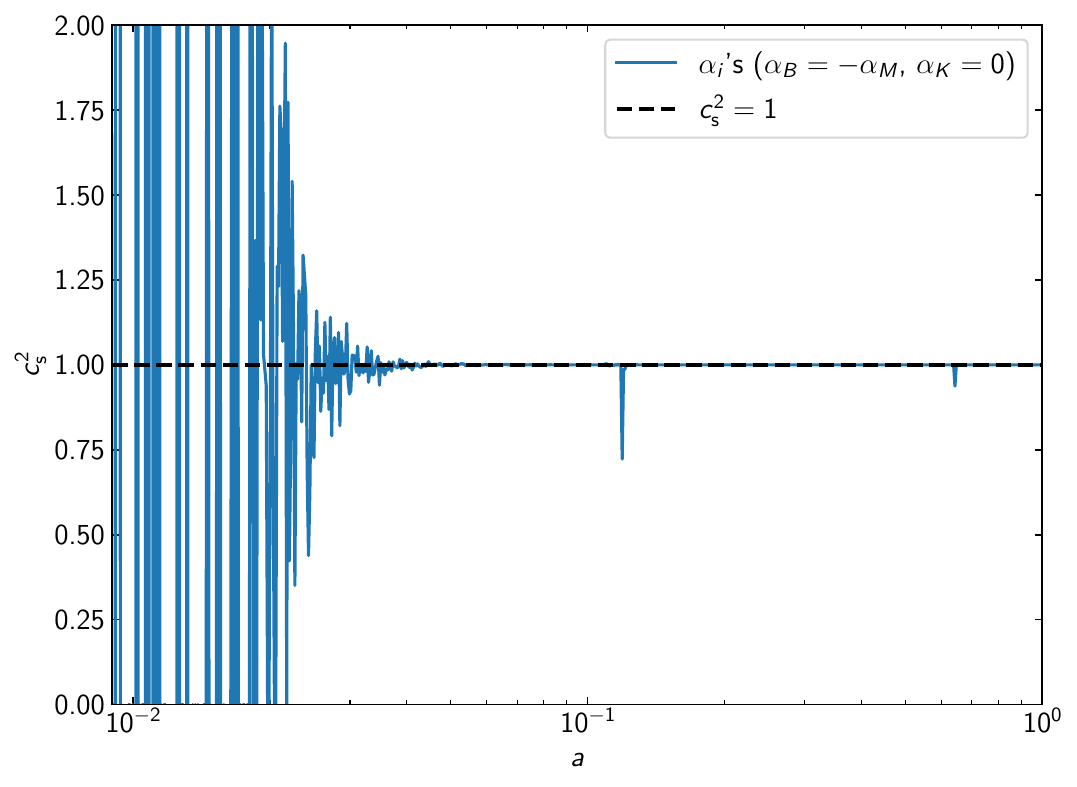}
    \end{minipage}%
    \begin{minipage}{.49\textwidth}
        \centering
        \includegraphics[width=\linewidth]{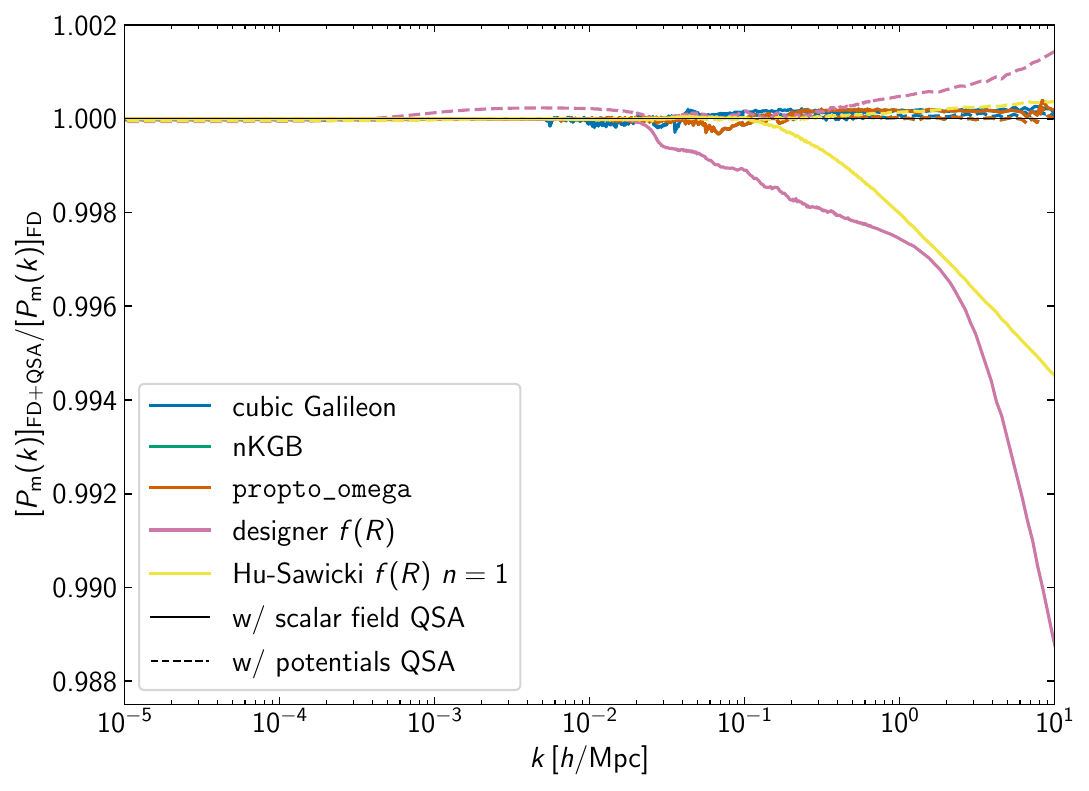}
    \end{minipage}
    \caption{\leftcap{} The speed of sound squared, \( c_s^2\), plotted against the scale factor, \( a \), for a designer \( f(R) \) gravity with \( w = -1 \) and \( B_0 = 0.01 \). The blue solid line illustrates the result derived from the Horndeski's \( \alpha_i \) functions, the Hubble parameter, and the background densities and pressures (Equation~\ref{eq:cs2}). The black dashed line indicates the expected theoretical value of \( c_s^2 = 1 \). The early-time oscillations in the derived quantity are the consequence of cancellation errors further amplified by the small magnitude of \( D_{\rm kin} \) in this model. These rapid variations and, more importantly, the violation of the stability condition \(c_s^2 > 0\) erroneously flag this viable model as unstable, motivating the need for an alternative parametrisation. \rightcap{} Effect of the QSA on the $z=0$ matter power spectrum for our test cosmologies. Each line represents the ratio of the matter power spectrum computed by selectively activating the QSA (see discussion in Section~\ref{sec:input}) to that obtained by following the full dynamics (FD) of the scalar field fluctuations. Solid lines indicate the QSA implemented in \hiclass{}, while dashed lines represent the same approach as implemented in \mgcamb{}. The two QSA strategies are generally consistent, although the methodology based on the metric potentials performs significantly better for the \fr{} gravity cosmologies on scales $k \gtrsim 0.1 \, \hMpc$.
    }
     \label{fig:cs2_and_qsa_hiclass}
\end{figure*}
One challenge when considering the entire class of Horndeski models is their potential for instability in response to perturbations. These instabilities result from unsuitable background solutions, and it is crucial to identify and exclude any parameter combinations leading to such unstable evolution. Physical instabilities come in three kinds~\citep[e.g.][]{Hu2013,Bellini2014}: (i) \textit{gradient instabilities} occur when the square of the speed of sound for perturbations turns negative as the background evolves. This negative value can cause perturbations to grow exponentially at small scales, destabilizing them over timescales that are comparable to the cutoff of the theory; (ii) \textit{ghost instabilities} arise when the sign of the kinetic term for background perturbations is incorrect. This issue is often considered in discussions of quantum stability. If ghost modes are present and dynamically interact, they can destabilize the vacuum, resulting in the production of both ghost and normal modes; (iii) \textit{tachyonic instabilities} manifest when the effective mass squared of scalar field 
perturbations is negative, leading to power-law instabilities at large scales growing faster than the Hubble scale. While the absence of ghost and gradient instabilities is an integral requirement for a healthy theory, tachyonic instabilities are not necessarily harmful. In fact, imaginary effective masses can still produce observationally viable models, and considering them as pathological {\it a priori} would amount to an overly conservative approach~\citep[see][for different treatments and interpretations of this instability]{Zumalacarregui2016,Frusciante2018,Gsponer2021}. Here, we will only impose that a theory must be free from ghost and gradient instabilities, which is expressed by the conditions
\begin{equation} \label{eq:stability_conditions}
    \Dkin > 0, \quad M_\ast^2 > 0, \quad \cs > 0 \, .
\end{equation}

Another class of instabilities that can plague Horndeski's models are the mathematical (or classical) instabilities~\citep{Hu2014}, which manifest as exponentially growing modes in the perturbations. The scalar field fluctuations, $V_X \equiv a\delta\phi/\phi^\prime$, follow the equation of motion
\begin{equation}\label{eq:vx_pert_schematic}
    \mathcal{A}(\tau)V_{X}^{\prime\prime}+ \mathcal{B}(\tau)V_{X}^{\prime} + \left[ \mathcal{C}(\tau) + k^2 \mathcal{D}(\tau) \right]V_{X} = \mathcal{E}(\tau,k) \, ,
\end{equation}
where the expressions for the time-dependent coefficients, $\{\mathcal{A},\dots,\mathcal{E}\}$, are detailed in Appendix~\ref{sec:useful_eqns}. To prevent significant growth of unstable modes over cosmological timescales, the following conditions must be satisfied:
\begin{itemize}
    \item If $\mathcal{B}^2(\tau) - 4\mathcal{A}(\tau)[ \mathcal{C}(\tau) + k^2\mathcal{D} (\tau) ] > 0$:
    \begin{equation}\label{eq:math_stability_1}
        \frac{-\mathcal{B}(\tau) \pm \sqrt{\mathcal{B}^2(\tau) - 4\mathcal{A}(\tau)[ \mathcal{C}(\tau) + k^2\mathcal{D} (\tau) ]}}{2\mathcal{A}(\tau)} < \xi H_0 \, .
    \end{equation}
    \item If $\mathcal{B}^2(\tau) - 4\mathcal{A}(\tau)[ \mathcal{C}(\tau) + k^2\mathcal{D} (\tau) ] < 0$:
    \begin{equation}\label{eq:math_stability_2}
        -\frac{\mathcal{B}(\tau)}{2\mathcal{A}(\tau)} < \xi H_0 \, .
    \end{equation}
\end{itemize}
Here, $\xi$ is a parameter controlling the level of instability in units of the Hubble constant, $H_0$, with larger values of $\xi$ allowing models to exhibit faster growing rates of these pathological modes.
 
For simple fixed-form parametrisations of the $\alpha$-functions (e.g., Equation~\ref{eq:propto_omega}), or even for theories based on a covariant formulation (e.g., Equations~\ref{eq:cubic_gal_lagrangian}-\ref{eq:fr_lagrangian}), finding stable models is just a matter of solving for the background evolution and check that the physical (and mathematical) conditions are satisfied. In these scenarios, we must only vary a handful of parameters and the search process is rather efficient. However, for more complex parametrisations producing a wider variety of functional forms, this trial-and-error approach can rapidly become sub-optimal. A mathematically equivalent, yet computationally more advantageous strategy consists in describing the Horndeski's Lagrangian at the level of background and linear perturbations using a basis that guarantees no-ghost and no-gradient instabilities from the start~\citep{Kennedy2018,Lombriser2018}. In practice, this can be achieved by replacing the $\alpha_i$'s with
\begin{equation}\label{eq:stable_basis}
    \Dkin, \quad M_\ast^2, \quad \cs, \quad \alphaBic \, ,
\end{equation}
together with a function describing the evolution of the background, such as the energy-density of the scalar field, $\rhophi$, its equation of state, $w_\phi$, or the Hubble parameter. The initial condition, $\alphaBic \equiv \alphaB(z=0)$, is used to integrate the non-linear differential equation for the braiding
\begin{equation}\label{eq:ode_braid}
    \dot\alpha_{B} + \left( 1 - \alphaM + \frac{\dot H}{H} \right)\alphaB - \frac{\alphaB^2}{2} = \mathcal{S}(x) \, ,
\end{equation}
where the source term is
\begin{equation}\label{eq:ode_braid_source}
    \mathcal{S}(x) \equiv -  2\alphaM + \csnum + \frac{\DeltaMpl}{M_\ast^2} \frac{2\dot H}{H} - \frac{3(\rhophi + \presphi)}{M_\ast^2 H^2} \, ,
\end{equation}
with the scalar field pressure, $\presphi$, obtained either from inverting the continuity equation (for more details see Section~\ref{sec:input}) or, if $w_\phi$ is provided, we must first integrate the continuity equation (see Appendix~\ref{sec:useful_eqns}) and then compute $\presphi = w_\phi\rhophi$. $\csnum \equiv \Dkin \cs$, $\DeltaMpl \equiv M_\ast^2 - 1$, and $\alphaM$ is given by Equation~\eqref{eq:run}. The Hubble parameter is obtained from the first Friedmann equation, $H = \rho_{\rm tot}$, where the total energy-density includes the contribution from the scalar field. Similarly, the second Friedmann equation gives $\dot H = -3(\rho_{\rm tot} + p_{\rm tot})/2H$. Note that Equation~\eqref{eq:ode_braid} follows from Equation~\eqref{eq:cs2} after expressing $H^\prime$ in terms of the density and pressure of the matter and scalar fields, and upon transforming conformal time derivatives into derivatives with respect to $\ln a$. Once a solution for $\alphaB$ is found, the kineticity, $\alphaK$, can be derived from Equation~\eqref{eq:Dkin}. 

It has been argued that stable basis functions offer no distinct advantages over the standard $\alpha$-functions in terms of computational efficiency or modeling capabilities \citep[e.g.,][]{Denissenya2018}. Contrary to this, our work suggests that reformulating Horndeski's gravity using the stable basis, Equation~\eqref{eq:stable_basis}, enhances our ability to efficiently sample viable theories and improves numerical stability. The left panel of Figure~\ref{fig:cs2_and_qsa_hiclass} illustrates how parametrisation with the $\alpha$-functions might misclassify models as unstable. Specifically, designer \fr{} gravity on a \lcdm{} background, which is inherently stable ($\cs=1$), shows rapid variations in the speed of sound due to inexact numerical cancellations when using Equation~\eqref{eq:cs2}, violating the no-gradient condition multiple times. In contrast, the stable basis directly incorporates the speed of sound as an input function, which, in this example, remains constant at unity throughout. Furthermore, using physically motivated parametrisations of the stable basis functions alongside highly optimised root finders and fast integrators (such as those implemented in modern Einstein-Boltzmann solvers) typically leads to rapid converge towards a solution for the braiding with the expected early-time evolution (see Section~\ref{sec:new_models} for details). Depending on the complexity of the parameter space, iteratively solving a differential equation is often more efficient than searching for a stable model starting from the $\alpha$-functions. 

\subsection{Quasi-static approximation}\label{sec:qsa}

The quasi-static approximation (QSA) is key to the study of cosmological perturbations within modified gravity and dark energy models. This approximation rests on the premise that the primary timescale of influence is the Hubble rate, and on scales smaller than the sound horizon ($ \cs k^2 \gg a^2 H^2 $) the time derivatives of the scalar field perturbations can be neglected owing to the rapid response of this degree of freedom to changes in the system. This simplifies the equations of motion for the scalar field into algebraic constraints, thus offering substantial computational benefits. In particular, for large values of the effective mass (i.e. the term proportional to $V_X$ in Equation~\ref{eq:vx_pert_schematic}) the scalar field perturbations undergo high-frequency oscillations that necessitate small integration steps. These can slow down computations significantly or even cause the solver to stall when the step size falls below machine precision. By employing the QSA, we can focus on the secular evolution of these perturbations, effectively excluding the high-frequency components.  

However, the QSA must be applied with care depending on the scale and the specific dynamics of the model. Einstein-Boltzmann solvers like \mgcamb{}~\citep{Hojjati2011,Zucca2019a,Wang2023} and \mgclass{}~\citep{Sakr2021} employ this approximation across all scales and throughout the entire evolution of the perturbations -- an approach that can potentially affect the accuracy of the predictions on large scales~\citep{Sawicki2015}. At the other end of the spectrum are solvers like \eftcamb{}~\citep{Hu2013} that avoid using the QSA entirely, a choice that can significantly degrade performance, in particular for certain sub-classes of models\footnote{See \cite{Hu2014} for the strategy adopted in \eftcamb{} to mitigate the impact of this choice.}. 
The \hiclass{} solver~\citep{Zumalacarregui2016,Bellini2019} adopts a hybrid strategy that leverages the computational efficiency of the QSA along with the accuracy provided by the full scalar field dynamics. Within this scheme, the QSA is selectively activated or deactivated for each time and scale individually, ensuring its use is optimized. The scalar field fluctuations and their time derivatives can be read off Equation~\eqref{eq:vx_pert_schematic} once the terms proportional to $V_X^{\prime\prime}$ and $V_X^{\prime}$ are neglected,
\begin{subequations}\label{eq:vx_hiclass_qsa}
    \begin{align}
        V_{X} &= \frac{\mathcal{E}(\tau,\,k)}{\mathcal{C}(\tau) + k^2 \mathcal{D}(\tau)}\,, \\
        V^\prime_{X} &= \frac{\diff}{\diff\tau} \left[ \frac{\mathcal{E}(\tau,\,k)}{\mathcal{C}(\tau) + k^2 \mathcal{D}(\tau)} \right]\,.
    \end{align}
\end{subequations}
It should be noted that, even within the quasi-static regime, $V_X$ remains time-dependent, implying that its time derivative does not vanish. Consequently, terms proportional to \( V^\prime_X \) (or time derivatives of other metric perturbations) in the Einstein equations, which are typically not reduced by small coefficients, must be accounted for in our calculations. 

Although the QSA implemented through Equation~\eqref{eq:vx_hiclass_qsa} provides a good description of the scalar field fluctuations and their derivatives, it can lead to small-scale inaccuracies in the matter perturbations for some models. This is illustrated in the right panel of Figure~\ref{fig:cs2_and_qsa_hiclass}, where the matter power spectrum for two of our test \fr{} gravity models (pink and yellow solid lines), computed by selectively activating the QSA in \hiclass{}, deviates by up to 1\% for \( k \gtrsim 0.1 \, h/\text{Mpc} \). For analyses restricted to the linear regime, these deviations are negligible. However, the linear power spectrum is also used as input for methods that predict non-linear structure formation~\citep[e.g.,][]{Bose2016, Aviles2017, Cusin2017, Cataneo2018, Giblin2019, Valogiannis2019}, and such small-scale inaccuracies can propagate to observables probing the non-linear regime, potentially exceeding the accuracy requirements established for Stage IV surveys~\citep[e.g.,][]{Taylor2018}.

A different QSA strategy consists in working directly with the modified metric potentials \citep{Bloomfield2012,Pace2020}. In the Newtonian gauge, the linearised modified Einstein field equation for the Newtonian potential, $\Psi$, and the intrinsic spatial curvature, $\Phi$, read \citep{Zhao2008,Hojjati2011}
\begin{subequations}\label{eq:potentials_qsa}
    \begin{align}
        k^{2}\Psi = -\frac{3}{2}\mu(a,k)a^{2}\left[\rhom\Delta_{\rm m}+3(\rhom+\presm)\sigma_{\rm m}\right] , \\
        k^{2}[\Phi-\gamma(a,k)\Psi] = \frac{9}{2}\mu(a,k)a^{2}(\rhom+\presm)\sigma_{\rm m} \, ,
    \end{align}
\end{subequations}
where the total comoving-gauge density perturbations are summed over all matter species $I \in \{ \text{c, b}, \text{r}, \nu, \dots \}$\footnote{Here and throughout we will be using the shorthand notation: c $\rightarrow$ cold dark matter, b $\rightarrow$ baryons, r $\rightarrow$ photons, $\nu \rightarrow$ neutrinos.}, $\rhom\Delta_{\rm m} \equiv \sum_I \rho_I\Delta_I$, with $\Delta_I$ being the comoving density contrast for the species $I$, and $(\rhom+\presm)\sigma_{\rm m} \equiv \sum_I (\rho_I+p_I)\sigma_I$ with $\sigma_I$ denoting the anisotropic stress of the individual matter fluids. Changes to the gravitational coupling experienced by non-relativistic particles are encoded in $\mu$, while the gravitational slip, $\gamma$, sources differences between the two metric potentials also in the absence of anisotropic stress. General Relativity is recovered for $\mu = \gamma = 1$. 

For the Horndeski's models described by the Lagrangian in Equation~\eqref{eq:scalar_horndeski} the two phenomenological functions take the simple forms \citep{Pogosian2016}
\begin{subequations}\label{eq:efe_qsa}
    \begin{align}
        \mu(k,a) = \frac{1}{M_\ast^2} \frac{\mu_{\rm p} + k^2 \csnum M_\ast^2 \mu_{\infty}/a^2 H^2}{\mu_{\rm p} + k^2 \csnum/a^2 H^2} \, , \\
        \gamma(k,a) = \frac{\mu_{\rm p} + k^2 \csnum M_\ast^2 \mu_{Z,\infty}/a^2 H^2}{\mu_{\rm p} + k^2 \csnum \mu_{\infty}/a^2 H^2} \, ,
    \end{align}
\end{subequations}
consistent with the notation of \cite{Pace2020}. Definitions for the quantities $\{ \mu_{\rm p}, \mu_{\infty}, \mu_{Z,\infty}\}$ are detailed in Appendix \ref{sec:useful_eqns}. The extension of \hiclass{} introduced in this work, \mochiclass{}, bypasses the QSA formulation from Equation~\eqref{eq:vx_hiclass_qsa} in favor of Equations~\eqref{eq:potentials_qsa} and~\eqref{eq:efe_qsa} in synchronous gauge, similar to \mgcamb{}\footnote{For the relevant equations, see \cite{Hojjati2011,Zucca2019a,Wang2023}. Here, we only report in Appendix \ref{sec:useful_eqns} a typo-corrected expression for the conformal time derivative of the scalar metric potential $\eta$.}. However, it activates the QSA selectively based on scale and time, when the scalar field's effective mass and the discrepancy between full and quasi-static solutions meet specific user-defined criteria~\citep[see][and Section~\ref{sec:code_description} for details]{Bellini2019}. The dashed lines in the right panel of Figure~\ref{fig:cs2_and_qsa_hiclass} demonstrate that the QSA approach implemented through the metric potentials is less affected by the neglected dynamics of the scalar field fluctuations compared to the implementation using Equation~\eqref{eq:vx_hiclass_qsa}.

\section{Code structure and implementation}\label{sec:code_description}
The code developed for this study, \mochiclass{}, builds on \hiclass{}, which itself extends the capabilities of the Einstein-Boltzmann solver \class{} \citep{Blas2011,Lesgourgues2011b} to evolve the background and linear perturbations of scalar-tensor theories within the Horndeski's framework. This section explores the new features introduced in \mochiclass{}. For a comprehensive review of the functionalities in \hiclass{} we refer the reader to \cite{Zumalacarregui2016, Bellini2019}.

\mochiclass{} enhances \hiclass{} by modifying the input, background, and perturbations modules to ensure greater numerical stability and facilitate a more comprehensive investigation of Horndeski's gravity. In the following, we provide a detailed description of these modifications.

\subsection{Input and precision parameters}\label{sec:input}

By setting the \texttt{gravity\_model} variable to \texttt{stable\_params}, users can specify a path to a file containing the functions $\DeltaMpl$, $\Dkin$, and $\cs$, as well as the $\ln a$ sampling times, using \texttt{smg\_file\_name}. Alternatively, these quantities can be directly assigned to the corresponding code variables: \texttt{Delta\_M2}, \texttt{D\_kin}, \texttt{cs2}, and \texttt{lna\_smg}. This direct assignment method is particularly recommended when running \mochiclass{} via the \texttt{classy} Python wrapper. To mitigate the impact of interpolation errors within the computational pipeline, we recommend using a time-sampling step size no larger than $\Delta\ln a = 7.5 \times 10^{-4}$ for the range $\ln a \in [-2,0]$. In \mochiclass{}, the \hiclass{} variable \texttt{parameters\_smg} serves a container for the initial condition, $\alphaBic$.

Users must also select an \texttt{expansion\_model} to define the background evolution of the scalar field. They can choose from the two \hiclass{} analytic expansion models, \{\texttt{lcdm}, \texttt{w0wa}\}, or opt for the new non-parametric models \{\texttt{wext}, \texttt{rho\_de}\}. In the \texttt{wext} model, the scalar field background density is integrated using the continuity equation (Equation~\ref{eq:continuity_phi}), with the background pressure defined as $\presphi = w_\phi \rhophi$. For the \texttt{rho\_de} model, the approach starts with the normalised background density $\normrhophi \equiv \rhophi/\rhophi(z=0)$, which is transformed to $\rhophi = \Omega_\phi H_0^2 \normrhophi$. Here, the Hubble constant, $H_0$, is an input parameter, and $\Omega_\phi = 1 - \sum_I \Omega_I$ represents the present-day normalised background energy-density in a flat universe. The scalar field background pressure in this configuration is then computed as
\begin{equation}\label{eq:pres_phi}
    \presphi = -\rhophi - \frac{\dot\rho_\phi}{3} \, .
\end{equation}

It is important to note that \hiclass{} employs a root-finding algorithm to determine the value of a user specified parameter necessary to achieve the present-day energy density, $\Omega_\phi H_0^2$ \citep[see Section 3.3 in][]{Bellini2019}. It can be the initial conditions of the scalar field, $\phi(\tau_{\rm ini})$, its time derivative, $\phi^\prime(\tau_{\rm ini})$, or any other parameter that contributes to the energy-density of the scalar field. This step is essential for models based on a covariant Lagrangian where $\rhophi$ is not known in advance, requiring the cosmological background to be solved self-consistently with the scalar field and the $\alpha$-functions. However, this becomes redundant in cases where the scalar field energy-density evolution is predefined, as often happens in the effective-theory approach. To optimize computational efficiency, \mochiclass{} has been modified to bypass the root-finding process. Instead, it directly uses $\Omega_\phi$ from the closure relation, $\Omega_{\rm tot} = 1$, to rescale the normalized energy density, $\normrhophi$.

Similarly to the stable basis functions, \texttt{wext} and \texttt{rho\_de} can read the necessary data either from a text file containing two columns \{$\ln a$, $w_\phi$ or $\normrhophi$\} located at a path specified in \texttt{expansion\_file\_name}, or directly from input arrays \texttt{lna\_de} and \texttt{de\_evo}. To minimise interpolation errors, it is recommended to use a time-sampling step-size $\Delta\ln a \leq 7.5 \times 10^{-4}$ for the range $\ln a \in [-2,0]$. In addition, when utilizing \mochiclass{}, users must activate the GR approximation scheme (see Section~\ref{sec:perturbations} below) by setting \texttt{method\_gr\_smg = on}. This feature allows the code to solve the standard system of equations down to a user-specified redshift, \texttt{z\_gr\_smg}. In practice, for redshifts greater than \texttt{z\_gr\_smg}, the scalar field behaves like a cosmological constant with $\rho_\Lambda = \rhophi(\texttt{z\_gr\_smg})$, and vanishing $\alpha_i$'s (see Equation~\ref{eq:lcdm_limit}). Given our focus on late-time modifications to GR, the default value for this transition epoch is set deep within the matter-dominated era, specifically at $\texttt{z\_gr\_smg} = 99$.

While the stable basis in \mochiclass{} ensures the positivity of the de-mixed kinetic term, the Planck mass, and the speed of sound, it does not automatically guarantee that the model is free from the $\alphaB = 2$ discontinuity (see Equation A.14 in~\cite{Zumalacarregui2016}), as this requires a solution for the braiding first. Therefore, we can safely set \texttt{skip\_stability\_tests\_smg = yes}, as \mochiclass{} always tests that the braiding never crosses 2, which was not implemented in \hiclass{}, and terminates returning a computation error~\footnote{This behavior, implemented at the level of the background module, acts as a safeguard to prevent code crashes during the calculation of the perturbations, specifically when the derivative of the metric perturbation, $h'$, is computed algebraically using Equation A.14 in \cite{Zumalacarregui2016}. However, this discontinuity can be entirely avoided by replacing the algebraic constraint with the differential equation for $h'$ derived from the trace of Einstein's field equations (Equation A.16 in \cite{Zumalacarregui2016}). In a future release of \mochiclass{}, we plan to implement a dynamic switch to the trace equation whenever an $\alphaB = 2$ crossing is detected.}. Furthermore, setting \texttt{skip\_math\_stability\_smg = no} allows the code to check for exponentially growing modes by enforcing the mathematical stability conditions outlined in Equations~\eqref{eq:math_stability_1} and \eqref{eq:math_stability_2}. Users can control the allowed instability growth rate via the \texttt{exp\_rate\_smg} variable, which defaults to 1.

Similar to \hiclass{}, \mochiclass{} has the option to treat the scalar field as a dynamical degree of freedom throughout the perturbations’ evolution, or to apply the QSA whenever is safe to do so\footnote{Unlike \hiclass{}, \mochiclass{} does not support enforcing QSA evolution at all times as \mgcamb{} does.}. These configurations can be selected by setting the \texttt{method\_qs\_smg} input variable to \texttt{fully\_dynamic} for the former, or \texttt{automatic} for the latter. In \texttt{automatic} mode, the QSA is activated for specific time intervals and wavenumbers based on whether the effective mass of the scalar field exceeds the threshold \texttt{trigger\_mass\_qs\_smg}, and the amplitude of scalar field fluctuations has reduced by a factor defined in \texttt{eps\_s\_qs\_smg}  \citep{Bellini2019}\footnote{\cite{Bellini2019} also introduce a radiation trigger to ensure that the scalar field mass is sufficiently large to counteract the oscillations inside the radiation sound horizon. While this requirement is relevant at early times, it becomes redundant at late times. Given that \mochiclass{} solves the standard GR equations throughout radiation domination and during the early stages of matter domination, this specific condition can be ignored here.}. Default values are set at \texttt{trigger\_mass\_qs\_smg = 100} and \texttt{eps\_s\_qs\_smg = 0.01}, which, even for large departures from the standard cosmology, affect the computed power spectra by $\lesssim 0.1\%$. \mochiclass{} can accurately solve the quasi-static equations down to $z=0$ (see Section~\ref{sec:qsa}), eliminating the need to enforce the full dynamic of the scalar field at low redshifts. Consequently, the corresponding control variable \texttt{z\_fd\_qs\_smg} can be safely set to 0.

\mochiclass{} also includes new precision parameters. Below is a list describing their scope and default values:

\begin{itemize}
    \item \texttt{background\_Nloga\_smg} and \texttt{loga\_split}: the background time-sampling strategy adopted for $a \gtrsim 0.05$ is crucial for achieving accurate interpolation of the quasi-static quantities $\{ \mu_{\rm p}, \mu_{\infty}, \mu_{Z,\infty}\}$. To this end, \mochiclass{} employs a denser sampling for $\ln a > \texttt{loga\_split}$, with the number of points controlled by \texttt{background\_Nloga\_smg}. For $\ln a < \texttt{loga\_split}$ the number of sampling points reverts to the standard set by the \class{} precision variable \texttt{background\_Nloga}. The default settings are \texttt{background\_Nloga = 3000}, \texttt{background\_Nloga\_smg = 3000}, and \texttt{loga\_split = -3}.

    \item \texttt{eps\_bw\_integration\_rho\_smg}: when using the \texttt{expansion\_model = wext}, where users define the equation of state for the scalar field, its background density is calculated from the continuity equation. This equation is integrated from \(a=1\) down to the scale factor \((1 - \texttt{eps\_bw\_integration\_rho\_smg})/(1+\texttt{z\_gr\_smg})\)\footnote{Note that this value must be larger than the smallest scale factor, \(a_{\rm min}\), in the input array \texttt{lna\_de}. We recommend setting \(\ln a_{\rm min} = -5\) for both \texttt{lna\_de} and \texttt{lna\_smg}.}. To ensure smooth derivatives of the scalar field pressure and related quantities at \texttt{z\_gr\_smg}, the default setting for \texttt{eps\_bw\_integration\_rho\_smg} is 0.1.

    \item \texttt{eps\_bw\_integration\_braid}: the equation for the braiding, Equation~\eqref{eq:ode_braid}, is integrated from $a=1$ backward to the scale factor \((1 - \texttt{eps\_bw\_integration\_braid})/(1+\texttt{z\_gr\_smg})\). To ensure smoothness at \texttt{z\_gr\_smg} for all functions derived from $\alphaB$ and its derivatives, we recommend setting \texttt{eps\_bw\_integration\_braid} to at least 0.1.

    \item \texttt{tol\_background\_bw\_integration}: this variable governs the integration accuracy of the continuity equation, Equation~\eqref{eq:continuity_phi}, and of the braiding equation, Equation~\eqref{eq:ode_braid}. Based on our analysis, setting \texttt{tol\_background\_bw\_integration = 1e-10} provides a good balance between computational speed and solution accuracy across all models examined in this study.

    \item \texttt{braid\_activation\_threshold}: Solving Equation~\eqref{eq:ode_braid} into the matter-dominated epoch poses numerical challenges, particularly for models that rapidly converge to GR at early times. While inaccuracies at high redshifts generally do not impact the late-time modified gravity phenomenology, they can introduce spurious instabilities that may incorrectly render a model non-viable. To minimize these effects, modifications to the Einstein equations are triggered only when the braiding solution reaches the \texttt{braid\_activation\_threshold}. Practically, this means adjusting the transition redshift, \texttt{z\_gr\_smg}, to match the redshift at which $|\alphaB|$ first equals \texttt{braid\_activation\_threshold}. Note that this threshold is not enforced for models with $|\alphaBic| \leq \texttt{braid\_activation\_threshold}$, e.g. quintessence and k-essence~\footnote{We can imagine fine-tuned models that, through precise adjustment of stable basis functions and the background, avoid early-time kinetic mixing between the scalar and metric degrees of freedom while still exhibiting a non-negligible running of the Planck mass (see, for example, Equation~\ref{eq:braid_early} below). To prevent introducing unwanted inaccuracies in the evolution of perturbations in these modified gravity models, we recommend setting \texttt{braid\_activation\_threshold} to 0, ensuring that \texttt{z\_gr\_smg} remains fixed at the original value specified by the user.}. The default value of \texttt{braid\_activation\_threshold = 1e-12} is chosen to ensure precise predictions across the broad range of deviations from GR examined in this study.

    \item \texttt{dtau\_start\_qs}: at the onset of modified gravity, marked by \texttt{z\_gr\_smg}, \mochiclass{} introduces the scalar field as a new degree of freedom by incorporating the equation for the field fluctuations (Equation~\ref{eq:vx_pert_schematic}) and its couplings to the metric perturbations into the Einstein-Boltzmann system. Initial conditions are defined by the QSA outlined in Equation~\eqref{eq:vx_hiclass_qsa}. The code then waits a conformal time specified by \texttt{dtau\_start\_qs} before assessing whether to activate the QSA, effectively turning the scalar field fluctuations into algebraic constraints. We found that \texttt{dtau\_start\_qs = 1e-7} works well for all models tested.
    
\end{itemize}

\subsection{Background}\label{sec:background}

Once the input arrays are prepared, the initial step involves calculating the background density of the scalar field. Depending on the chosen expansion model, the code performs one of the following actions: (i) integrates the continuity equation and stores both $\rhophi$ and $\presphi$ for later use; (ii) rescales the input normalized background energy-density of the scalar field to compute $\rhophi$, and calculates $\presphi$ using Equation~\eqref{eq:pres_phi}, saving both values for future reference; (iii) analytically determines the scalar field density and pressure, along with other background quantities, as already implemented in \hiclass{}.

Subsequently, with access to the stable basis functions, the scalar field density and pressure, as well as $H$ and $\dot H$, the code solves for $\alphaB$ by backward integrating Equation~\eqref{eq:ode_braid} from its present-day value, $\alphaBic$. It calculates the kineticity, $\alphaK$, using Equation~\eqref{eq:Dkin}, and if necessary updates \texttt{z\_gr\_smg} based on the \texttt{braid\_activation\_threshold}. However, it is important to acknowledge a slight inconsistency in our approach: $\rhophi$, $\presphi$, and associated variables are not adjusted to emulate a cosmological constant over the interval between the updated and the original \texttt{z\_gr\_smg} values. This methodological choice is made to save computational resources while accounting for the very small effect of the modified background on the perturbations. Within the Horndeski's framework we expect changes to the growth of structure to be linked to changes to the background expansion. Therefore, viable models exhibiting significant departures from standard gravity only in recent epochs, presumably also closely match a \lcdm{} expansion history until recently.   

Finally, with all relevant background functions now available, \mochiclass{} computes the quasi-static quantities $\{ \mu_{\rm p}, \mu_{\infty}, \mu_{Z,\infty}\}$ and their derivatives. These are also stored in the background table and forwarded to the perturbations module for further processing.

\subsection{Perturbations}\label{sec:perturbations}

The evolution of the perturbations closely follows that implemented in \hiclass{}, except for the new GR approximation scheme and the effective field equations used for the QSA. While \hiclass{} follows the dynamics of the scalar field and its coupling to the metric tensor already prior to recombination, \mochiclass{} activates the additional degree of freedom only after the time \texttt{z\_gr\_smg}. This is achieved by enabling \texttt{method\_gr\_smg}, which directs the solver to apply the standard \class{} equations up to the transition redshift. 

Once the GR approximation is deactivated, the initial conditions for the scalar field are derived from the quasi-static expressions given in Equation~\eqref{eq:vx_hiclass_qsa}, transitioning from the standard \class{} equations to those of \hiclass{}. This approximation scheme is crucial to manage the discontinuity at \texttt{z\_gr\_smg}, which otherwise leads to computational instability by forcing the integrator to attempt reductions in the step size to below double precision levels. Consistent with other approximation schemes used, the initial conditions for the metric and stress-energy tensor perturbations are set from the integration time step immediately preceding the deactivation.

With the initial conditions now defined, \mochiclass{} can use the QSA scheme of \hiclass{} for the evolution of the perturbations. Should the scalar field effective mass and fluctuations amplitude satisfy specific criteria~\citep{Bellini2019}, the QSA may be selectively activated or deactivated up to six times. However, within \mochiclass{}, this scheme is only applicable to each wavenumber, $k$, following a designated time, $\tau(\texttt{z\_gr\_smg}) +  \texttt{dtau\_start\_qs}$. Furthermore, rather than applying the QSA directly to the scalar field fluctuations, $V_X$, \mochiclass{} targets the metric potentials, thereby enhancing numerical accuracy. Accordingly, the quasi-static part of \hiclass{} has been restructured to align with the methodology developed for \mgcamb{}.

\section{Code validation and performance}\label{sec:code_validation}

\begin{figure*}[ht]
    \centering
    \includegraphics[width=\textwidth]{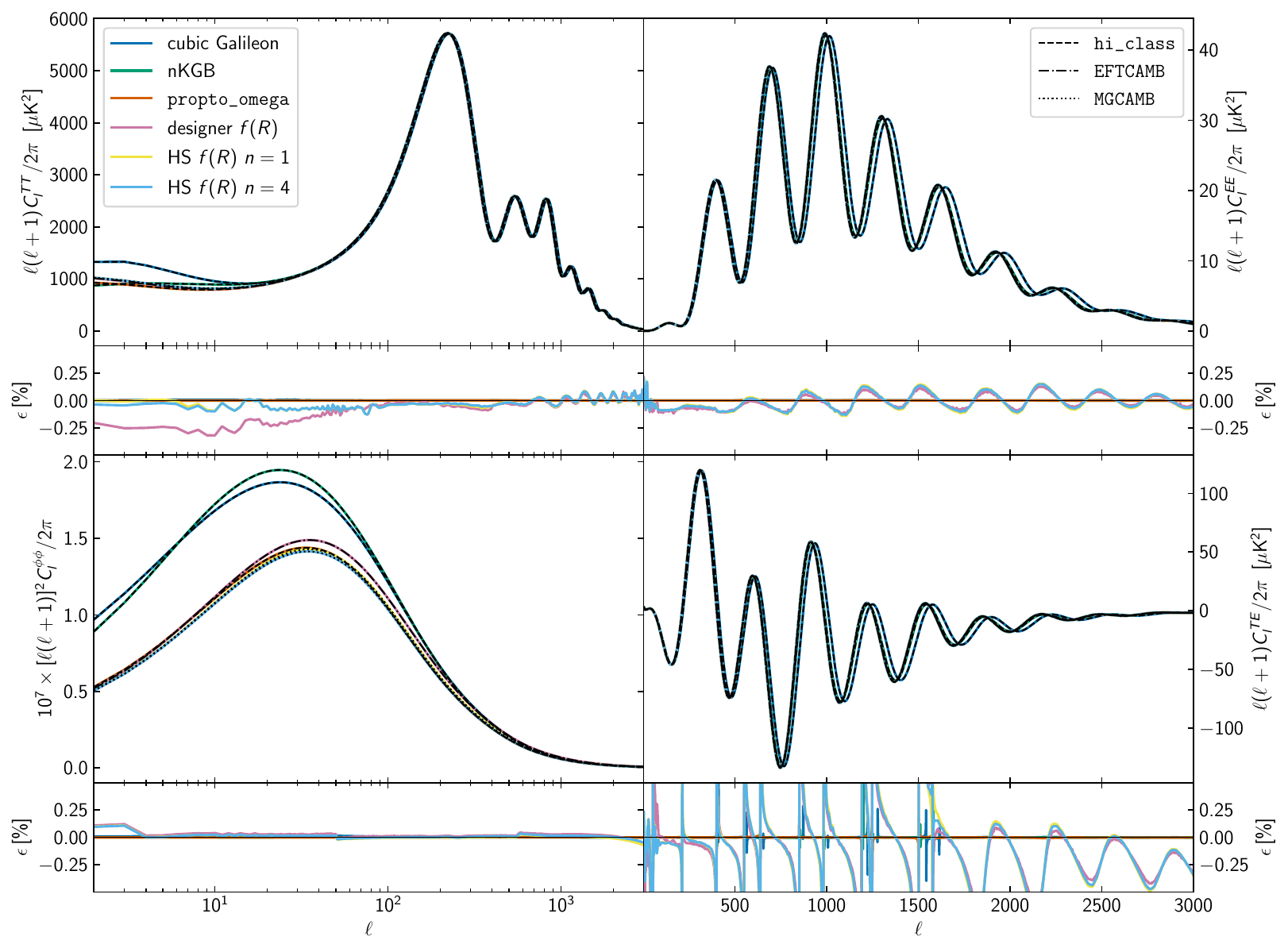}
    \caption{Comparison of the CMB auto- and cross-power spectra as computed by \mochiclass{} (colored lines) against predictions from well-validated Einstein-Boltzmann solvers (black lines). The dashed lines for cubic Galileon and nKGB models, as well as \(\proptoomega{}\), are generated by \hiclass{} following the full scalar field dynamics. For designer \( f(R) \) gravity, the dot-dashed lines are obtained with \eftcamb{}, while the dotted lines for Hu-Sawicki (HS) \( f(R) \) models are produced by \mgcamb{} using the quasi-static approximation. The fractional differences, \( \epsilon \), shown in the lower sub-panels quantify the agreement between \mochiclass{} in \texttt{automatic} mode and the reference spectra obtained with the other Einstein-Boltzmann solvers. Notably, all discrepancies are below 0.25\% except in the TE cross-spectra, where the reference spectra’s zero crossings amplify the deviations. All discrepancies larger than 0.05\% for the \( f(R) \) models stem primarily from differences between \class{} and \camb{}, and are also present in \lcdm{} (not shown).}
    \label{fig:mochi_v_xcamb_hiclass_cmb}
\end{figure*}

\begin{figure*}[ht]
    \centering
    \includegraphics[width=\textwidth]{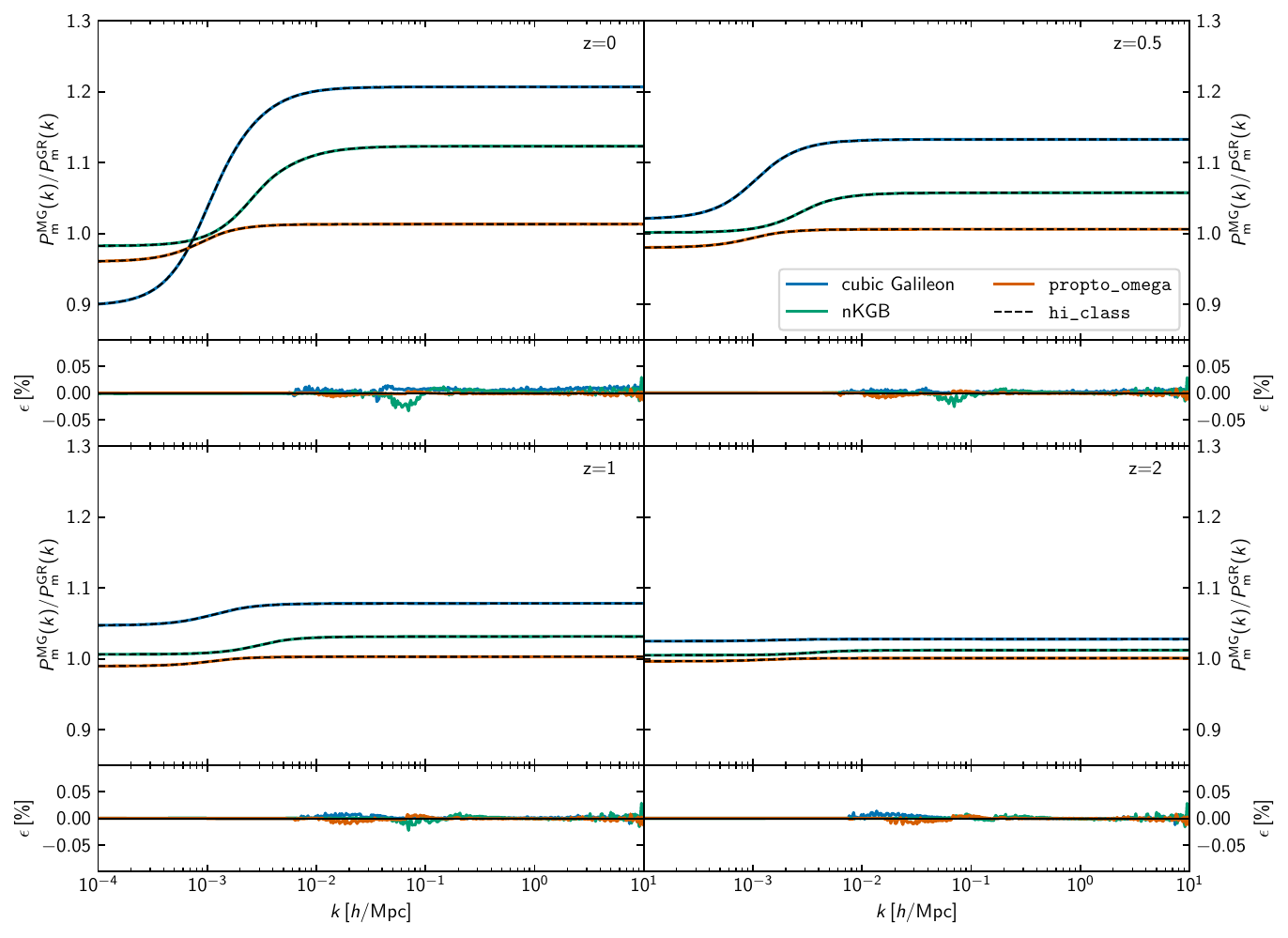}
    \caption{\mochiclass{} accuracy quantification for the matter power spectrum through comparison with \texttt{hi\_class} across various redshifts (\(z = 0, 0.5, 1, 2\)). Each main panel displays the matter power spectrum ratio in modified gravity models relative to \lcdm{} as calculated by \mochiclass{} (colored lines) and \hiclass{} (dashed lines). The sub-panels show the fractional deviation of \mochiclass{} from \hiclass{}, indicating an agreement within 0.05\% for the considered redshifts. This level of accuracy demonstrates the robustness of the computational strategy adopted in \mochiclass{}, which includes an alternative parametrisation of the Horndeski's functions, a QSA formulated in terms of modifications to the equations for the metric potentials, and the activation of modified gravity deep into the matter-dominated era.}
    \label{fig:mochi_v_hiclass_pk}
\end{figure*}

\begin{figure*}[ht]
    \centering
    \includegraphics[width=\textwidth]{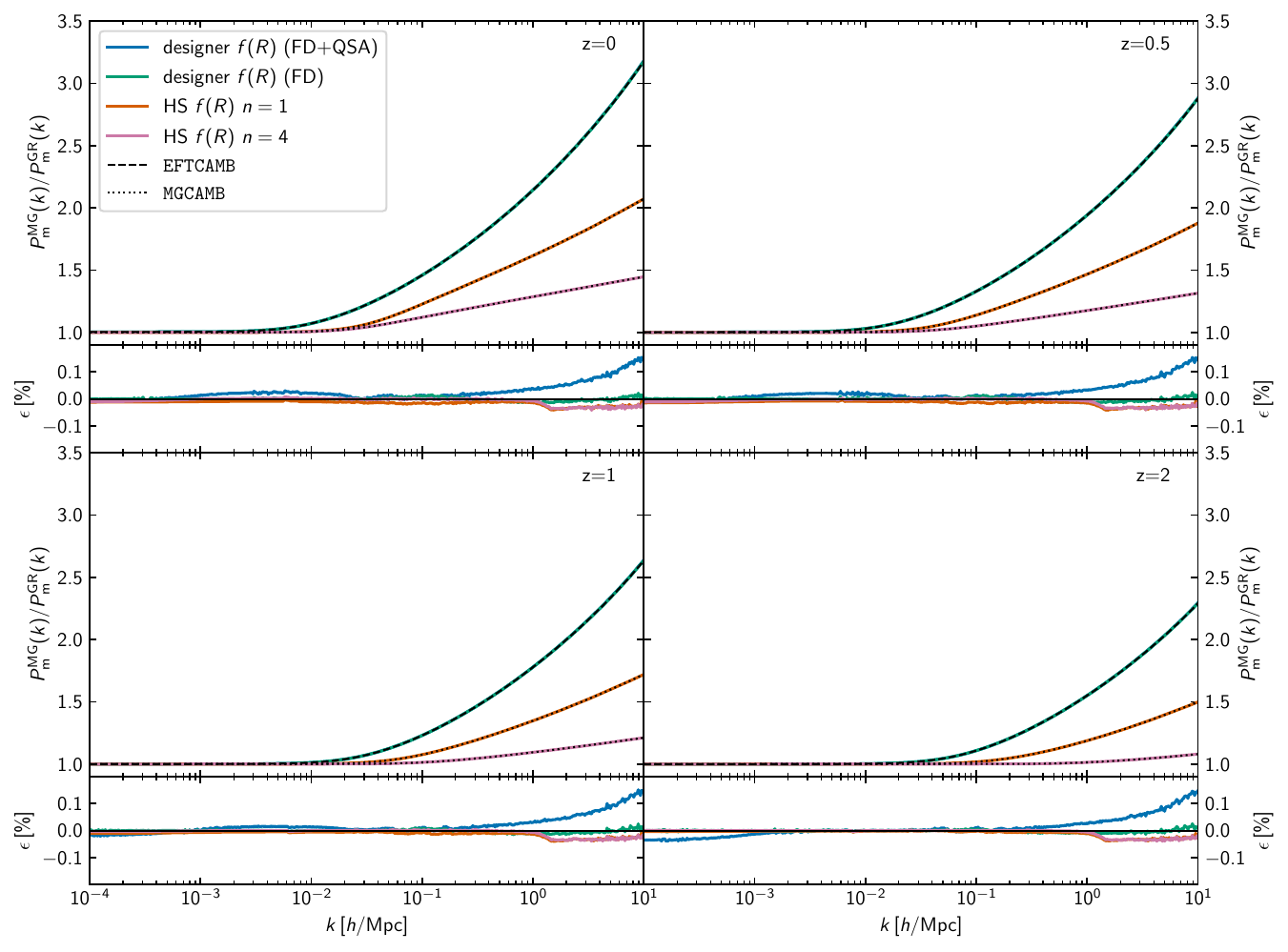}
    \caption{Extending Figure~\ref{fig:mochi_v_hiclass_pk}, here we consider modified gravity models characterised by a scale-dependent growth on sub-horizon scales. The reference power spectra for \fr{} gravity are computed with \eftcamb{} (dashed) and \mgcamb{} (dotted). The lower panels focus on the fractional deviation of \mochiclass{} from the \texttt{CAMB}-based solvers, providing a direct comparison for the matter power spectrum ratios shown above. The Hu-Sawicki (HS) models demonstrate a close match, with deviations from \mgcamb{} remaining under 0.025\% across all scales. In contrast, for the designer \fr{} model, \mochiclass{} shows a consistent discrepancy that increases at smaller scales (blue line), a result predominantly due to the quasi-static approximation. This difference, in fact, significantly reduces when \mochiclass{} always integrates the full scalar field dynamics (green line).}
    \label{fig:mochi_v_xcamb_pk}
\end{figure*}

Before exploring the new capabilities of \mochiclass{}, we first compare its predictions with those from the established codes \hiclass{}~\citep{Zumalacarregui2016,Bellini2019}, \eftcamb{}~\citep{Hu2013,Hu2014}, and \mgcamb{}~\citep{Hojjati2011,Zucca2019a,Wang2023}. This comparative analysis allows for a thorough assessment of all test models and ensures the robustness of our results across different parametrisations, approximations, and integration strategies. Although not discussed here, similar agreements between \mochiclass{} and the other Einstein-Boltzmann solvers are achieved when including massive neutrinos. This is especially true for the small, non-factorisable contributions to the growth of structure caused by their coupling to modified gravity.

\subsection{CMB spectra}

Figure~\ref{fig:mochi_v_xcamb_hiclass_cmb} displays the CMB multipoles for temperature (upper left), polarization (upper right), their cross-correlation (lower right), and the lensing potential power spectrum (lower left). The lower sub-panels contain the relative deviations of the \mochiclass{} predictions (coloured lines) compared to the results from other Einstein-Boltzmann solvers (black lines). As reference spectra, we use \hiclass{} without the QSA for cubic Galileon, nKGB, and \proptoomega{} models; \eftcamb{} for designer \fr{} gravity; and \mgcamb{} for Hu-Sawicki \fr{} gravity. These quantities were generated using the precision settings detailed in Appendix~\ref{sec:precision} for both \camb{} and \class{}.

First, it is important to note that the differences between \mochiclass{} and \hiclass{} are barely visible at this scale. This suggests that the newly implemented GR approximation at early times, the reparametrisation using stable basis functions, and the application of the quasi-static approximation until late in the perturbation evolution, all introduce minimal numerical errors.

The agreement between \mochiclass{} and the \camb{}-based codes is very good, with differences remaining well within $\sim 0.25\%$. Further analysis reveals that mismatches exceeding 0.05\% come from variations between \class{} and \camb{}, similarly affecting \lcdm{} predictions (not shown). This validates the QSA implementation in \mochiclass{}, which closely aligns with \mgcamb{}, and confirms the reliability of the QSA/full dynamics switching feature in \hiclass{}. Note that despite \eftcamb{} and \mgcamb{} both deriving from \camb{}, discrepancies around 0.2\% are observed between these codes for the low-$\ell$ temperature multipoles. Initially, these could be attributed to the activation of the QSA in \mochiclass{}, potentially affecting the accuracy of the integrated Sachs-Wolfe (ISW) effect calculations, particularly in the designer \fr{} model where deviations from GR are more pronounced. However, further investigation indicated that these differences are due to the use of different \camb{} versions, with \mgcamb{} employing a more recent release, and also persist for the standard cosmology (not shown).

\subsection{Matter power spectrum}

Figure~\ref{fig:mochi_v_hiclass_pk} presents the evolution of the matter power spectrum ratio, $P_{\rm m}^{\rm MG}(k)/P_{\rm m}^{\rm GR}(k)$, for the three cosmologies characterised by scale-independent growth on sub-horizon scales. The lower sub-panels show that for these models \mochiclass{} (coloured lines) deviates by an average of approximately $0.01\%$ from \hiclass{} (black lines). Differences are negligible for scales $k \lesssim 0.01 \, \hMpc$, where the QSA is never activated. 

To isolate the impact of the new \mochiclass{} features, Figure~\ref{fig:mochi_v_xcamb_pk} also shows the matter power spectrum of the \fr{} gravity predictions relative to same quantity in \lcdm{}. For the Hu-Sawicki models the agreement between \mochiclass{} (coloured lines) and \mgcamb{} (dotted lines) is excellent across all redshifts. Differences larger that 0.2\% are visible only for $k \gtrsim 1 \, \hMpc$, regardless of model and time. Using an independent \texttt{Mathematica} notebook employing the QSA, we computed the linear growth functions both in HS $f(R)$ gravity and in \lcdm{}~\cite[see, e.g., Eq. 81 in][]{Brax2013}, and by taking their ratio we verified that these small differences originate from \mgcamb{}.

Moreover, Figure~\ref{fig:mochi_v_xcamb_pk} illustrates the (albeit small) inaccuracies introduced by the QSA in the designer model. The blue lines, generated using a hybrid method that activates the QSA under certain conditions (detailed in Section~\ref{sec:code_description}), should be compared with the \eftcamb{} predictions (dashed lines), which do not employ the QSA. As expected, the small scales are most affected by this approximation, due to the conditions for its activation being met for more extended time intervals. To verify that deviations from \eftcamb{} are caused by the QSA, we forced \mochiclass{} to follow the full dynamics of the scalar field (green line). The lower sub-panels confirm that \mochiclass{} predictions now align with \eftcamb{} at a level of \(10^{-4}\).

\subsection{Performance}\label{sec:performance}

{
\renewcommand{\arraystretch}{1.5} 
\begin{table*}[ht]
    \centering
    \begin{tabular}{c c c c c c}
    \toprule
                    & input (I) & background (B) & perturbations (P) & (I)+(B)+(P) & total  \\
    \midrule
    \midrule
    Cubic Galileon  & -80\% (0.02)   & +100\% (0.025) & -52\% (9.5) & -52\% (9.6) & -40\% (13.2)   \\ 
    \proptoomega{}  & -75\% (0.012)  & +100\% (0.016) & -25\% (9.9) & -24\% (9.9) & -17\% (13.3)  \\
    \bottomrule
    \end{tabular}
    \caption{Variations in execution time for the input (I), background (B), and perturbations (P) modules in \mochiclass{}, relative to \hiclass{}. The values in parentheses are the \mochiclass{}'s execution times in seconds averaged over 10 single-thread runs on an Apple M1 Pro chip. Both codes used the \texttt{automatic} mode, with \hiclass{} forcing full dynamical evolution of the scalar field from $z=10$, as recommended by its developers. The models considered—Cubic Galileon, based on the covariant Lagrangian, and \proptoomega{}, based on effective theory—are representative of the two approaches implemented in \hiclass{}. The column labeled (I)+(B)+(P) lists the cumulative variations across the three modules, emphasising that the evolution of the perturbations is the primary bottleneck. The final column indicates that, overall, enabling the QSA at low redshifts can lead to substantial performance improvements, with minimal impact on accuracy (see Figure~\ref{fig:cs2_and_qsa_hiclass}). For a detailed comparison between the execution time with and without QSA see Fig. 11 in~\cite{Bellini2019}.
    }
    \label{tab:performance}
\end{table*}
}

Here we assess the impact of the \mochiclass{} patch on execution time, examining both individual modules and the overall runtime. Table~\ref{tab:performance} summarises the relative changes in execution time compared to \hiclass{} for the input, background, and perturbations modules, as well as their combined effect and the total runtime variation, including the other unmodified modules. To ensure a fair comparison, we evaluate two models that activate different features in \hiclass{}: the cubic Galileon model, which requires preparatory steps to iteratively determine suitable initial conditions and map the Horndeski $G_{i}$'s to the $\alpha$-basis, and the \proptoomega{} model, which involves direct analytic expressions for both the $\alpha$'s and the scalar field background evolution.

Contrary to intuition, \mochiclass{} shows reduced runtime for both models at the input level, despite \proptoomega{} only needing to fill the relevant arrays. This is because \hiclass{} unnecessarily applies the root finder for the initial conditions to this model--a step not performed by \mochiclass{}. Unsurprisingly, the execution of the background module takes twice as long as in \hiclass{}, due to the integration of the differential equation for the braiding function. However, this slowdown is less severe than suggested by~\cite{Denissenya2018}. Assuming a uniform scanning strategy of their two-dimensional parameter space and a stable region amounting to 22\% of the prior volume, the number of unstable models is 3.5 times that of the viable models. Therefore, starting from the stable basis ensures an overall 56\% speedup for the background module compared to using the $\alpha$-functions in this particular scenario. This efficiency gap becomes even wider for complex parametrisations, such as those discussed in Section~\ref{sec:new_models}, where stable models can be particularly challenging to find.

The third column of Table~\ref{tab:performance} highlights the advantage of enabling the QSA for $z<10$: it can reduce the computing time for the evolution of perturbations by up to 50\%. Since this part of the code contributes most significantly to the cumulative execution time of the three modified modules (fourth column), any performance improvement here directly impacts the total runtime of the Einstein-Boltzmann solver (last column). The difference between the two codes arises from enabling  the activation of the QSA in \mochiclass{} throughout the perturbation evolution, while in \hiclass{} we apply it only up to some high redshift (i.e., $z=10$), as recommended by its developers.

\section{Extended Parametrisation for Horndeski gravity}\label{sec:new_models}
\begin{figure}[ht]
    \centering
    \includegraphics[width=0.49\textwidth]{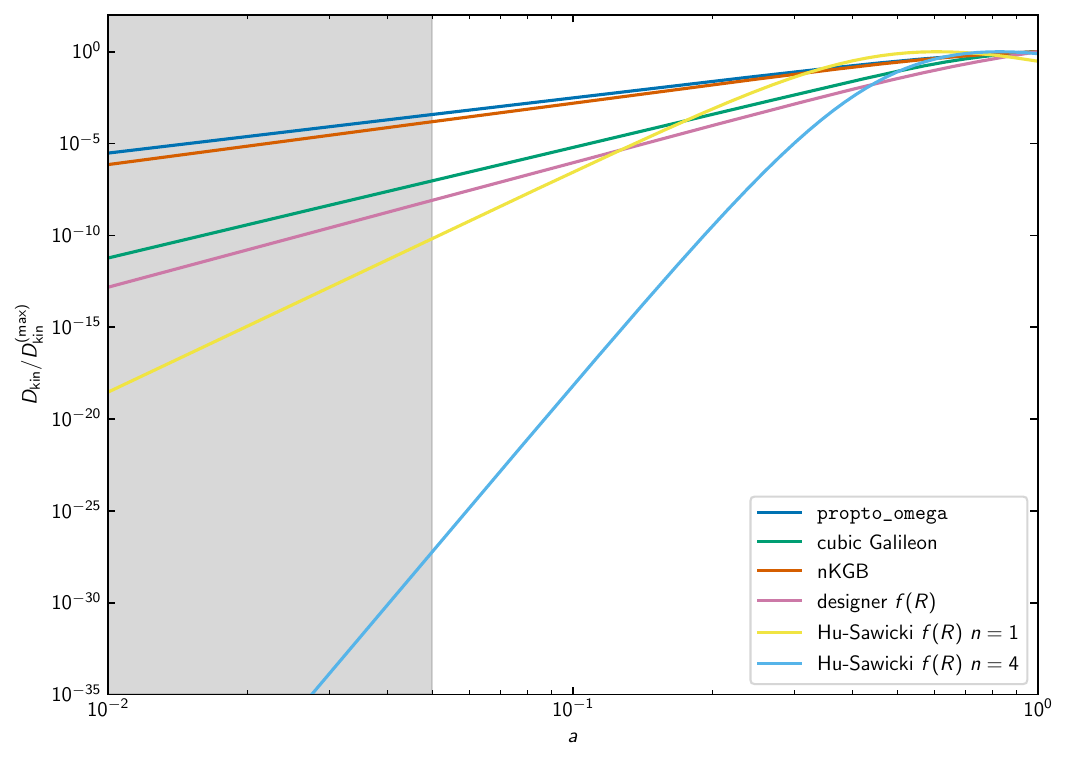}
    \caption{Evolution of the de-mixed kinetic term, \( \Dkin \), normalised to its maximum value, \( \Dkin^{\rm (max)} \), across a variety of modified gravity theories. At sufficiently early time (shaded area) \( D_{\text{kin}} \) closely follows a power-law behavior for all models. A similar behavior (not shown) characterises departures from the cosmological strength of gravity, \( \DeltaMpl \), and the background energy density of the scalar field, \( \rhophi \).}
    \label{fig:Dkin_all_models}
\end{figure}
After establishing \mochiclass{} accuracy and efficiency using extensively studied modified gravity cosmologies, we can now leverage its new features to input general functions of time and explore the phenomenology of Horndeski gravity in greater detail. For this purpose, a parametrisation that generalises \proptoomega{} and includes all covariant theories analysed in Section~\ref{sec:code_description} could be extremely valuable~\citep{Linder2016}.

A common feature of all our test models is that, deep in the matter-dominated era, their stable basis functions either approach the GR limit as power laws, $A_J a^{\zeta_J}$, or tend to constant values, $C_J$, to a very good approximation. For instance, Figure~\ref{fig:Dkin_all_models} illustrates that, regardless of the specifics governing the dynamics of the scalar field, the de-mixed kinetic term is well described by a power law for $a \lesssim 0.05$ (shaded area). Therefore, any late-time evolution of the stable functions can be interpreted as deviations, $\Delta_J$, from this early-time power-law behavior. Specifically, we can express the stable functions as
\begin{subequations}\label{eq:new_parametrisation}
    \begin{align}
        \DeltaMpl &= \sign(A_M) e^{\zeta_M \cdot x + b_M + \Delta_M(x)} \, , \\
        \Dkin &= e^{\zeta_D \cdot x + b_D + \Delta_D(x)} \, , \\
        \cs &= \left[ \sign(A_{c_{\rm s}}) e^{\zeta_{c_{\rm s}} \cdot x + b_{c_{\rm s}}} + C_{c_{\rm s}} \right] e^{\Delta_{c_{\rm s}}(x)} \, , \\
        \normrhophi &= \left[ \sign(A_\rho)e^{\zeta_\rho \cdot x + b_\rho} + C_\rho \right] \Delta_\rho(x) \, ,
    \end{align}
\end{subequations}
where $x \equiv \ln a$, $b_J \equiv \ln |A_J|$, $\sign$ denotes the sign function, and for $x \ll 0$ we have the limiting values
\begin{subequations}\label{eq:deviation_limits}
    \begin{align}
        \Delta_M, \Delta_D, \Delta_{c_{\rm s}} &\rightarrow 0 \, , \\
        \Delta_{\rho} &\rightarrow 1 \, .
    \end{align}
\end{subequations}
The constant $C_\rho$ is not a free parameter; it is determined by the constraint $\normrhophi(x=0) = 1$. For negative $A_M$ and $A_{c_{\rm s}}$, we must ensure that $\DeltaMpl > -1$ and $\cs > 0$ throughout the evolution. Given that the power-law term for the speed of sound generally represents a small correction to the constant $C_{c_{\rm s}}$ (as discussed below), it can be neglected when generating new models. Therefore, the gradient-free condition implies simply that $C_{c_{\rm s}} > 0$.

\subsection{Parametrising departures from constants and power laws}\label{sec:gp_pca}

\begin{figure*}[ht]
    \centering
    \begin{minipage}{.49\textwidth}
        \centering
        \includegraphics[width=\linewidth]{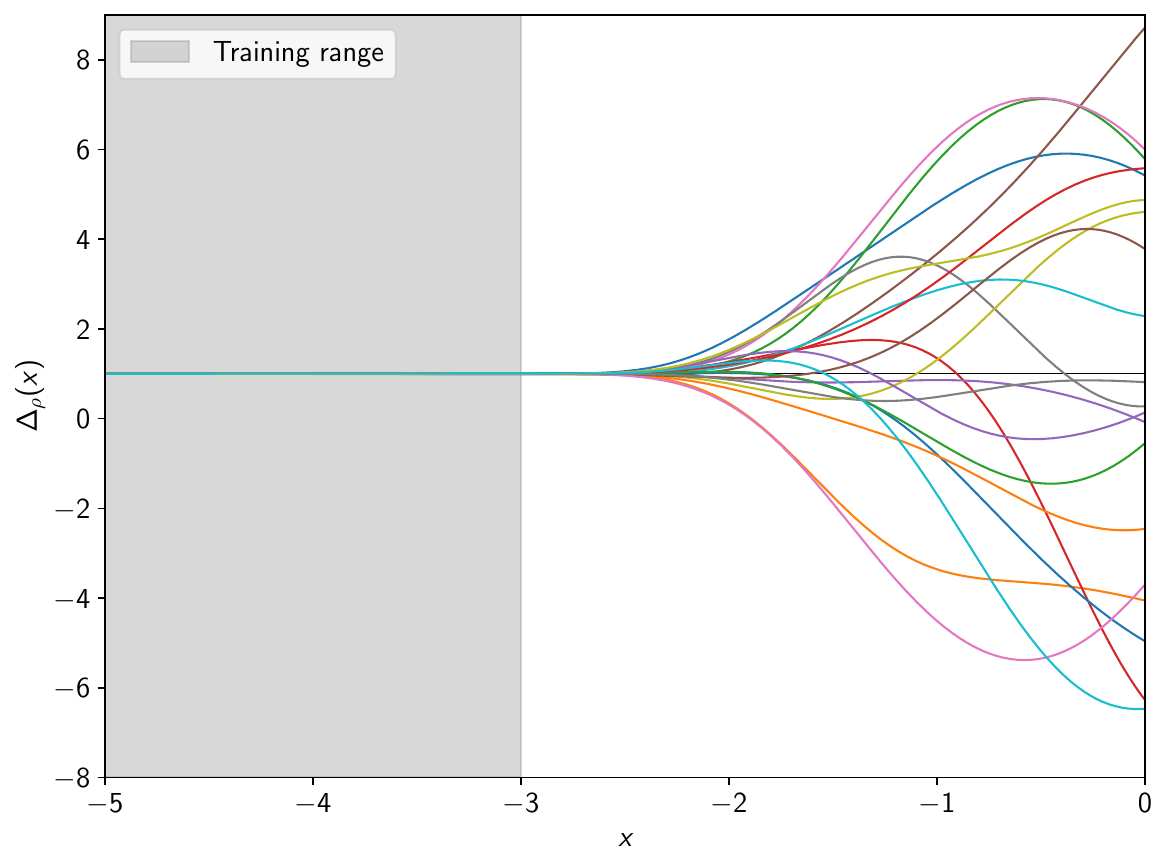}
    \end{minipage}%
    \begin{minipage}{.49\textwidth}
        \centering
        \includegraphics[width=\linewidth]{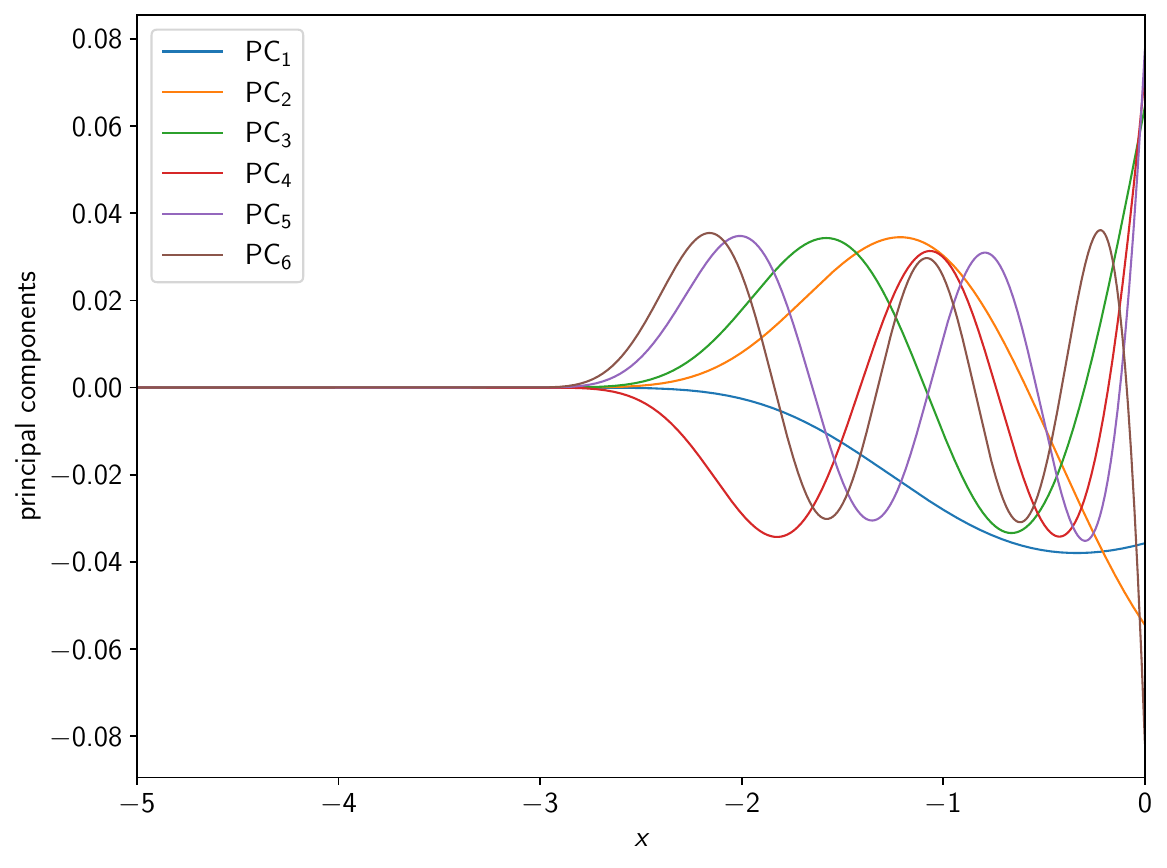}
    \end{minipage}
    \caption{\leftcap{} Samples from a Gaussian Process (GP) representing the correction factor, \( \Delta_{\rho} \), which quantifies deviations in the scalar field background energy-density from a baseline power-law behavior as a function of time, \( x \equiv \ln a \). The GP's hyper-parameters are detailed in the main text, and its mean is set to 1. The shaded region marks the training range, over which the GP is constrained to unity, coinciding with the interval where the background energy-density of the scalar field closely follows a power-law behaviour. The various trajectories of the 20 curves outside the training range represent possible extrapolations of \( \Delta_{\rho} \) beyond the power-law regime. \rightcap{} The first six principal components (PCs) extracted from 10,000 samples generated by the Gaussian Process described in the left panel. The PCs illustrated here are sufficient to reconstruct the scalar field background energy-density of all test models to better than 1\%, thus demonstrating the efficiency of the dimensionality reduction achieved by the principal component analysis (PCA). Similar performance apply to PCAs for the other input functions within the stable basis.}
    \label{fig:rnd_samples_and_pcs}
\end{figure*}

Next, we need to find a sufficiently general yet simple parametrisation for the functions $\Delta_J$. A commonly employed strategy to parametrise the EFT functions or the $\alpha_i$'s involves expanding them as ratios of polynomials, i.e., Pad\'e approximants~\citep[see, e.g.,][]{Gleyzes2017,Peirone2017}. Here, we opt for an alternative methodology grounded in Gaussian Processes (GPs)~\citep[see, e.g.,][for detailed discussions]{Rasmussen2006,Aigrain2022} and Principal Component Analysis (PCA) applied to the deviations $\Delta_J$. In broad terms, we proceed as follows:
\begin{enumerate}
    \item We generate training data such that the GP is consistent with Equation~\eqref{eq:deviation_limits}. Practically, this requires an array of zeroes over the interval $x \in [-5,-3]$.
    
    \item After choosing a GP kernel, $K(x_i,x_j)$, for a particular basis function, and values for its hyperparameters, we let the GP learn from the training data. After the training, the GP posterior collapses to a delta function centered around zero over the range $x \in [-5,-3]$, while the prior increasingly dominates the more we move into the untrained region. Note that all our GPs have zero mean.

    \item We draw thousands of samples from the trained GP, calculate their mean, and subtract it from each sample so that the transformed samples have zero mean.

    \item We perform PCA on the transformed samples and retain the first $N_J$ principal components. 

    \item We project the specific de-trended basis function (i.e. we first remove the constant and/or power-law contributions in Equation~\ref{eq:new_parametrisation}) of each test model (cubic Galileon, nKGB, etc.) onto the selected principal components and quantify the accuracy of the reconstructed approximation.

    \item For each basis function, we independently iterate over steps 2-5 to identify the kernel, hyperparameter values, and number of principal components that achieve percent-level reconstruction accuracy for $x > -3$ across all test models simultaneously.
\end{enumerate}

It is reasonable to assume that a covariant theory of gravity in the Horndeski class should be described by basis functions that vary smoothly with time. This assumption restricts the choice of the GP kernel to either the squared exponential form, 
\begin{equation}\label{eq:SEK}
K(x_i,x_j) = \sigma^2 \exp\left[ -\frac{(x_i - x_j)^2}{2\ell^2} \right] \, , 
\end{equation}
or the rational quadratic form,
\begin{equation}\label{eq:RQK}
K(x_i,x_j) = \sigma^2\left[1 + \frac{(x_i - x_j)^2}{2\vartheta\ell^2} \right]^{-\vartheta} \, . 
\end{equation}
Here, the variance $\sigma^2$ represents the average distance of the generated samples away from the mean, the lengthscale $\ell$ determines the oscillatory features of the samples, and the parameter $\vartheta$ in the rational quadratic kernel controls the relative weighting of large-scale and small-scale variations~\footnote{Note that the squared exponential kernel follows from the rational quadratic kernel in the limit $\vartheta \rightarrow +\infty$.}. Intuitively, we expect that the most relevant hyperparameters controlling the functional forms generated by the GP are $\ell$ and $\vartheta$, as $\sigma^2$ acts as a simple rescaling factor~\footnote{In other words, the principal components are insensitive to variations in $\sigma^2$.}. Therefore, we fix $\sigma^2 = 10$, and only vary $\ell$ and $\vartheta$ together with the number of principal components to achieve our target reconstruction accuracy of 1\%. 
As an example, the left panel of Figure~\ref{fig:rnd_samples_and_pcs} displays random functions, $\Delta_\rho$, sampled from a trained GP using the rational quadratic kernel, as described by Equation~\eqref{eq:RQK}, with $\ell = 1$ and $\vartheta = 5$. The right panel of the same figure shows the first six principal components derived from a total set of 10,000 samples, collectively accounting for 90\% of the cumulative explained variance, which meets the desired accuracy threshold~\footnote{Note that although the length-scale parameter of the GPs is $\sim 1$, the principal components can exhibit variations on much shorter time scales. By adjusting the relative importance of the principal component weights, one can generate functional forms that evolve more rapidly than what the GP's length-scale parameter would typically permit.}. Similar considerations apply to the other basis functions; Table~\ref{tab:gp_pca} details the kernel, hyperparameter values, and number of principal components used for each. To keep interpolation errors in \mochiclass{} under control, our time-sampling strategy for the random curves (and consequently for the principal components) closely follows that described in Section~\ref{sec:input}. Specifically, we use 3,000 sampling points for the range $[-3,0]$ and 200 sampling points for the range $[-5,-3]$.
{
\renewcommand{\arraystretch}{1.5} 
\begin{table}[ht]
    \centering
    \begin{tabular}{c c c c c}
    \toprule
                    & Kernel & $\ell$ & $\vartheta$ & \# PCs \\
    \midrule      
    \midrule
    $\Delta_M$  & RQ  & 1.0 & 3.0 & 7  \\ 
    $\Delta_D$  & SE  & 1.5 & -- & 6 \\
    $\Delta_{c_{\rm s}}$  & RQ & 1.0 & 4.0 & 6 \\ 
    $\Delta_\rho$  & RQ  & 1.0 & 5.0  & 6 \\
    \bottomrule
    \end{tabular}
    \caption{Kernel form (rational quadratic (RQ) or  squared exponential  (SE)), corresponding hyperparameters, and number of retained principal components (\# PCs) used to model each of the deviation functions, $\Delta_{J}$, in Equation~\eqref{eq:new_parametrisation}.}
    \label{tab:gp_pca}
\end{table}
}

\subsection{Reconstruction of test models}\label{sec:reconstruction}

\begin{figure*}[ht]
    \centering
    \begin{minipage}{.49\textwidth}
        \centering
        \includegraphics[width=\linewidth]{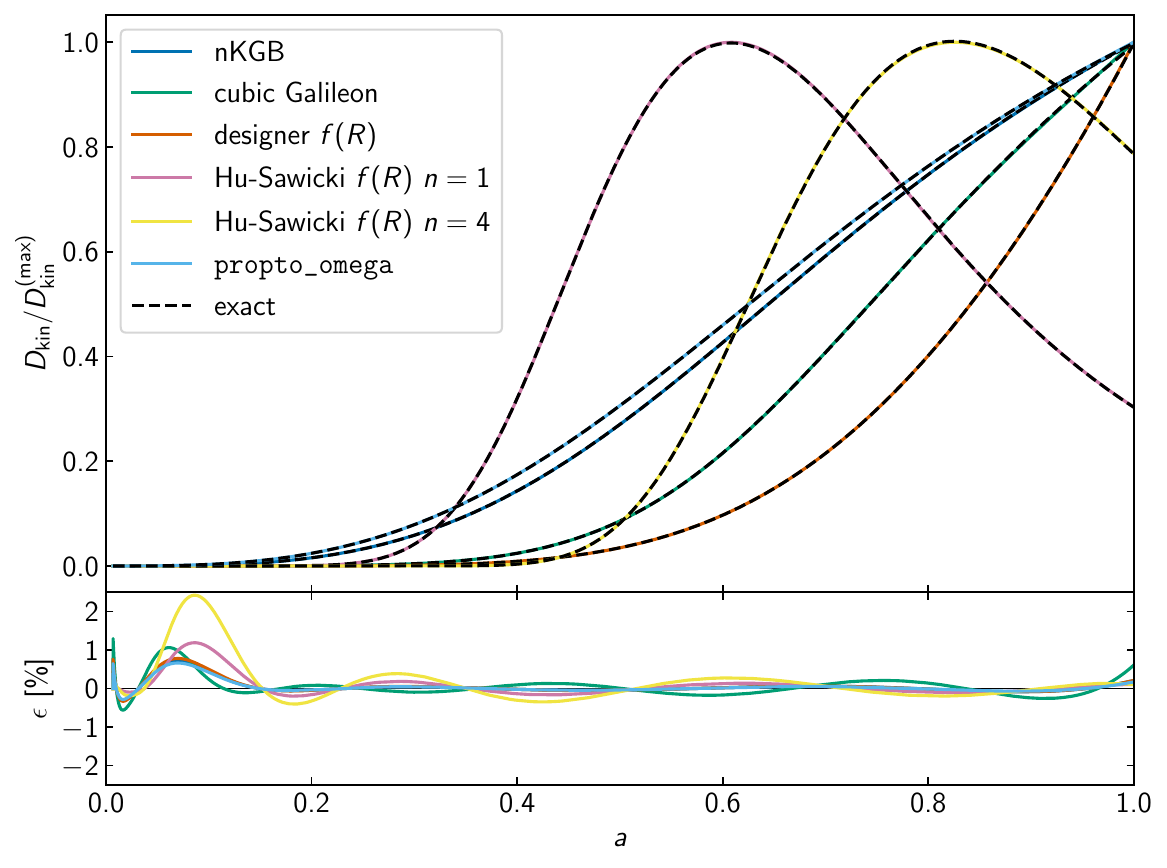}
    \end{minipage}%
    \begin{minipage}{.49\textwidth}
        \centering
        \includegraphics[width=\linewidth]{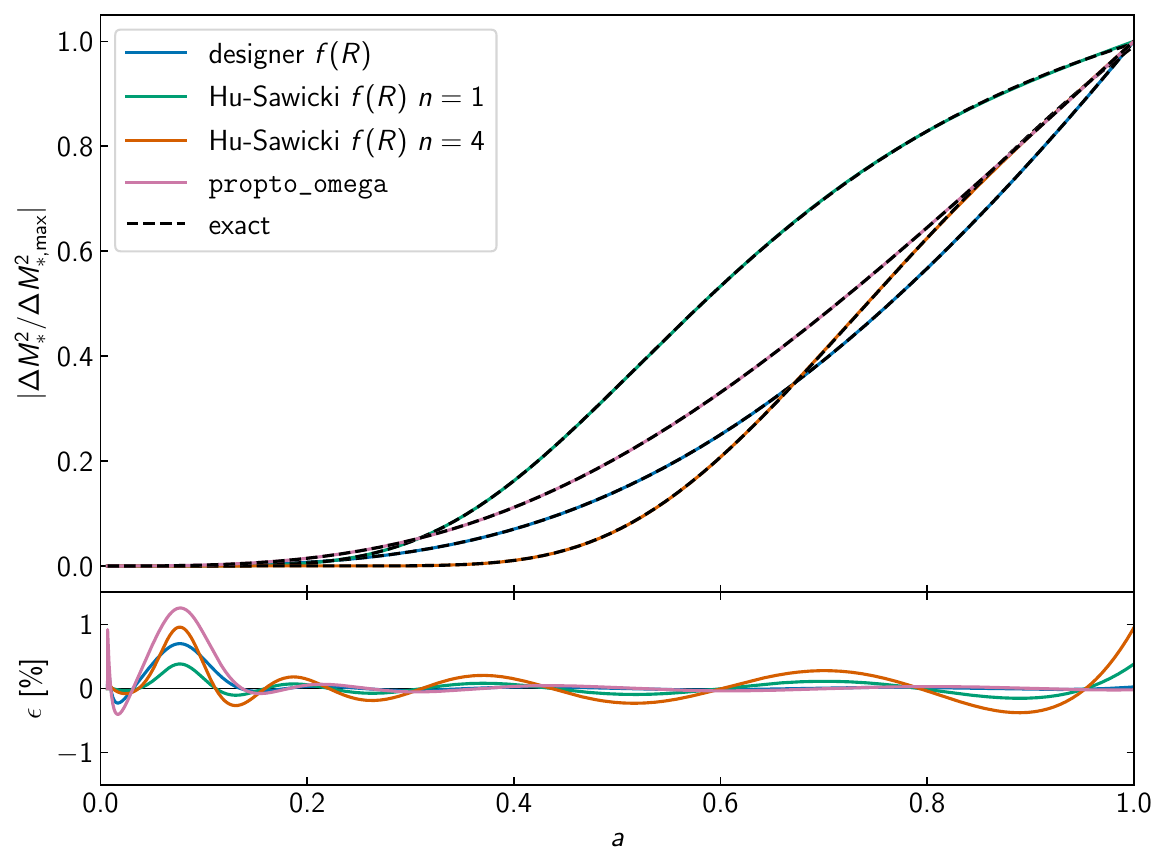}
    \end{minipage}
    \caption{Reconstruction of the stable basis functions, with the left panel displaying the normalized de-mixed kinetic term (\(D_{\text{kin}}/D_{\text{kin}}^{\rm (max)}\)), and the right panel showing the normalized absolute deviations from the standard cosmological strength of gravity, \( |\DeltaMpl/\Delta M_{\ast,{\rm max}}^2| \). Coloured lines indicate the functions as reconstructed from the principal components and power-law approximation, while the dashed lines are the exact quantities. The lower sub-panels quantify the reconstruction accuracy, which is consistently better than 1\% for the entire range where modified gravity significantly affect the growth of structure \(a > 0.1\).}
    \label{fig:Dkin_DeltaM_reconstruction}
\end{figure*}

\begin{figure}[ht]
    \centering
    \includegraphics[width=0.49\textwidth]{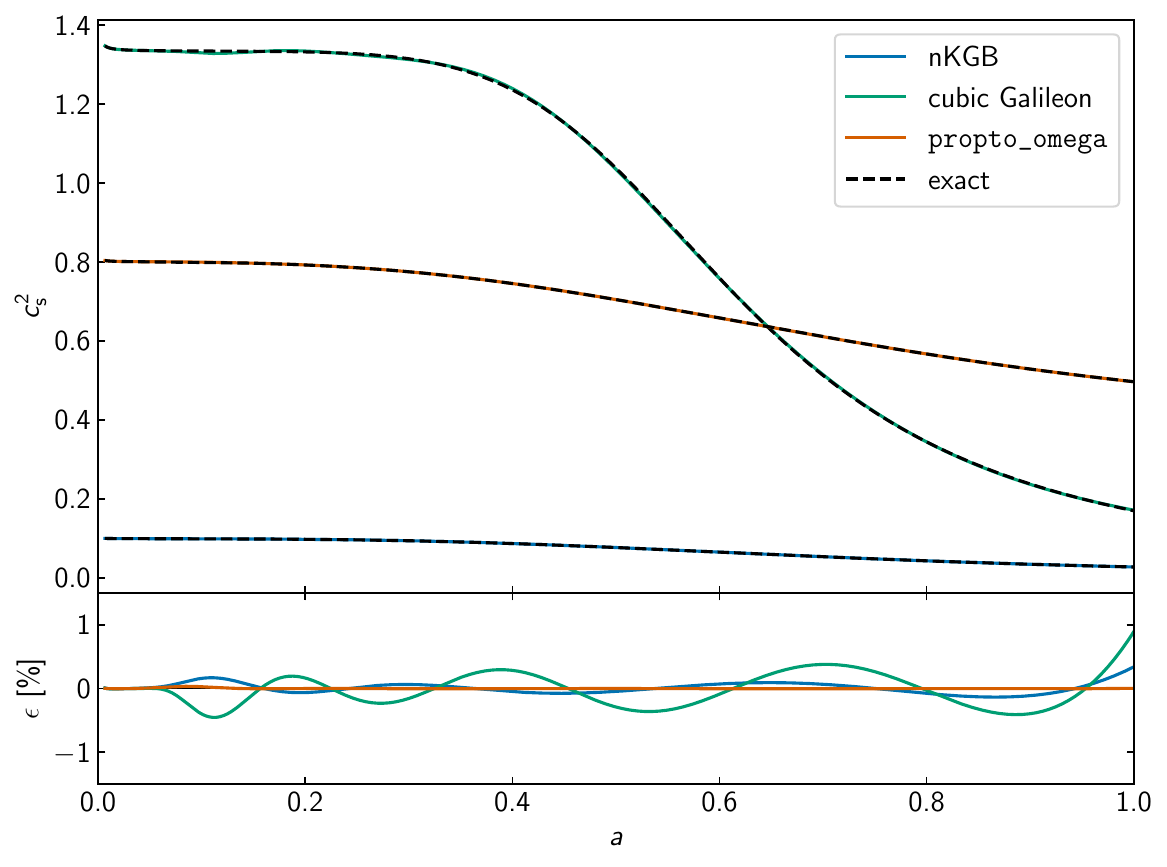}
    \caption{Reconstructed (coloured) and exact (dashed) speed of sound for the test models exhibiting non-trivial evolution (i.e., \( \cs \neq 1 \)). The lower panel demonstrates that the PCA-based reconstruction can achieve better than 1\% accuracy at all times.} 
    \label{fig:cs2_reconstruction}
\end{figure}

\begin{figure*}[ht]
    \centering
    \begin{minipage}{.49\textwidth}
        \centering
        \includegraphics[width=\linewidth]{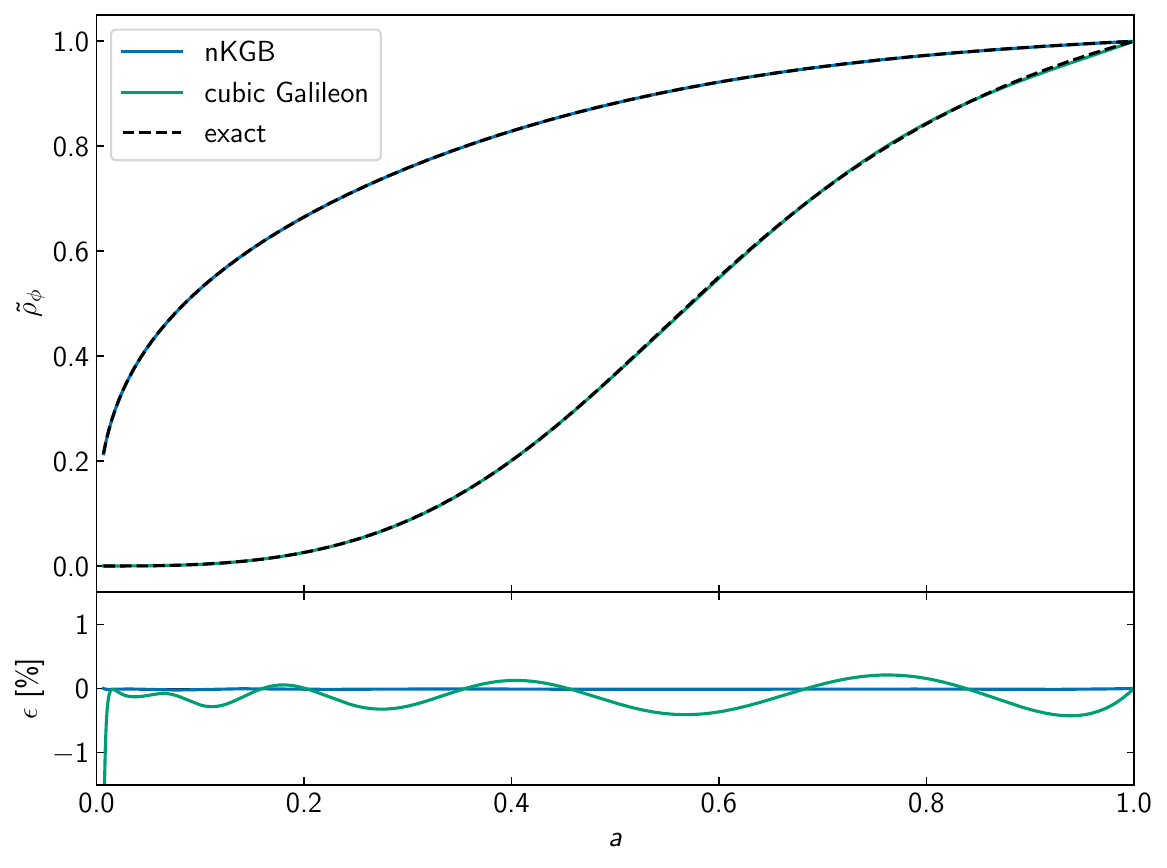}
    \end{minipage}%
    \begin{minipage}{.49\textwidth}
        \centering
        \includegraphics[width=\linewidth]{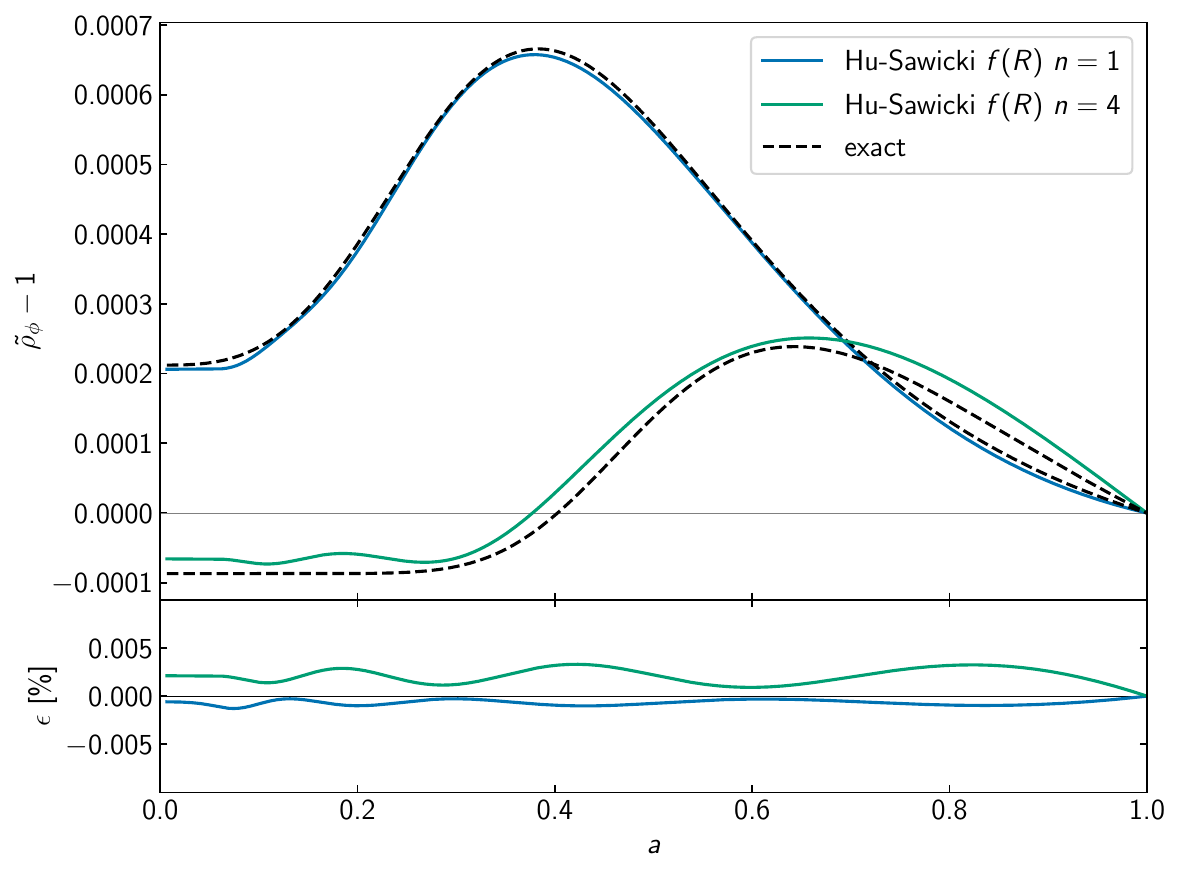}
    \end{minipage}
    \caption{Comparison between the reconstructed background energy-density of the scalar field (coloured lines) and the exact calculations (dashed lines). The left panel shows the evolution of the energy-density for two models minimally coupled to gravity, whereas the right panel displays \( \rhophi \) for two variants of the Hu-Sawicki (HS) \fr{} gravity model which feature a conformal coupling to gravity. For both types of models, the reconstructed quantities closely align with the exact references, maintaining the accuracy within 0.5\% for scale factors \( a > 0.02 \).}
    \label{fig:rho_phi_reconstruction}  
\end{figure*}

\begin{figure*}[ht]
    \centering
    \includegraphics[width=\textwidth]{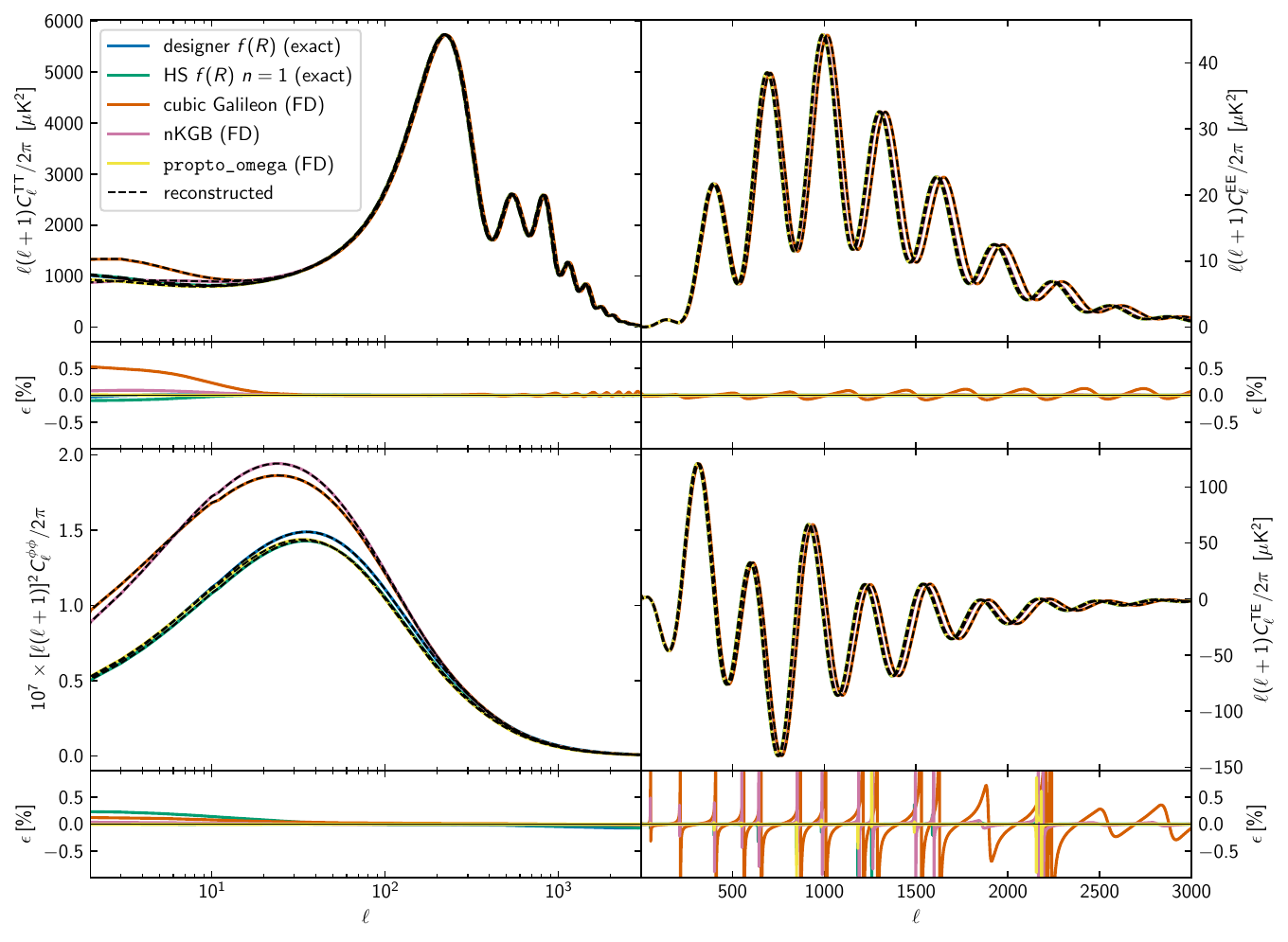}
    \caption{Comparison of the CMB auto- and cross-power spectra obtained with \mochiclass{} using the reconstructed stable basis (dashed lines) against the exact predictions (coloured lines). For the cubic Galileon and nKGB models, as well as \(\proptoomega{}\), the reference spectra are generated by \hiclass{} following the full scalar field dynamics (FD). The exact calculations for the two \fr{} gravity models are performed using \mochiclass{} with the \texttt{automatic} method and precisely determined stable basis functions. For all reconstructed models, the initial conditions for the braiding parameter, \( \alphaBic \), are optimised to minimise the matter power spectrum's relative deviation at \( z=0 \) compared to the exact models (refer to Equation~\ref{eq:mse}). The small fractional differences, \( \epsilon \), shown in the lower sub-panels indicate that the reconstructed models are observationally indistinguishable from their exact counterparts.}
    \label{fig:cmb_reconstruction}
\end{figure*}

\begin{figure*}[ht]
    \centering
    \includegraphics[width=\textwidth]{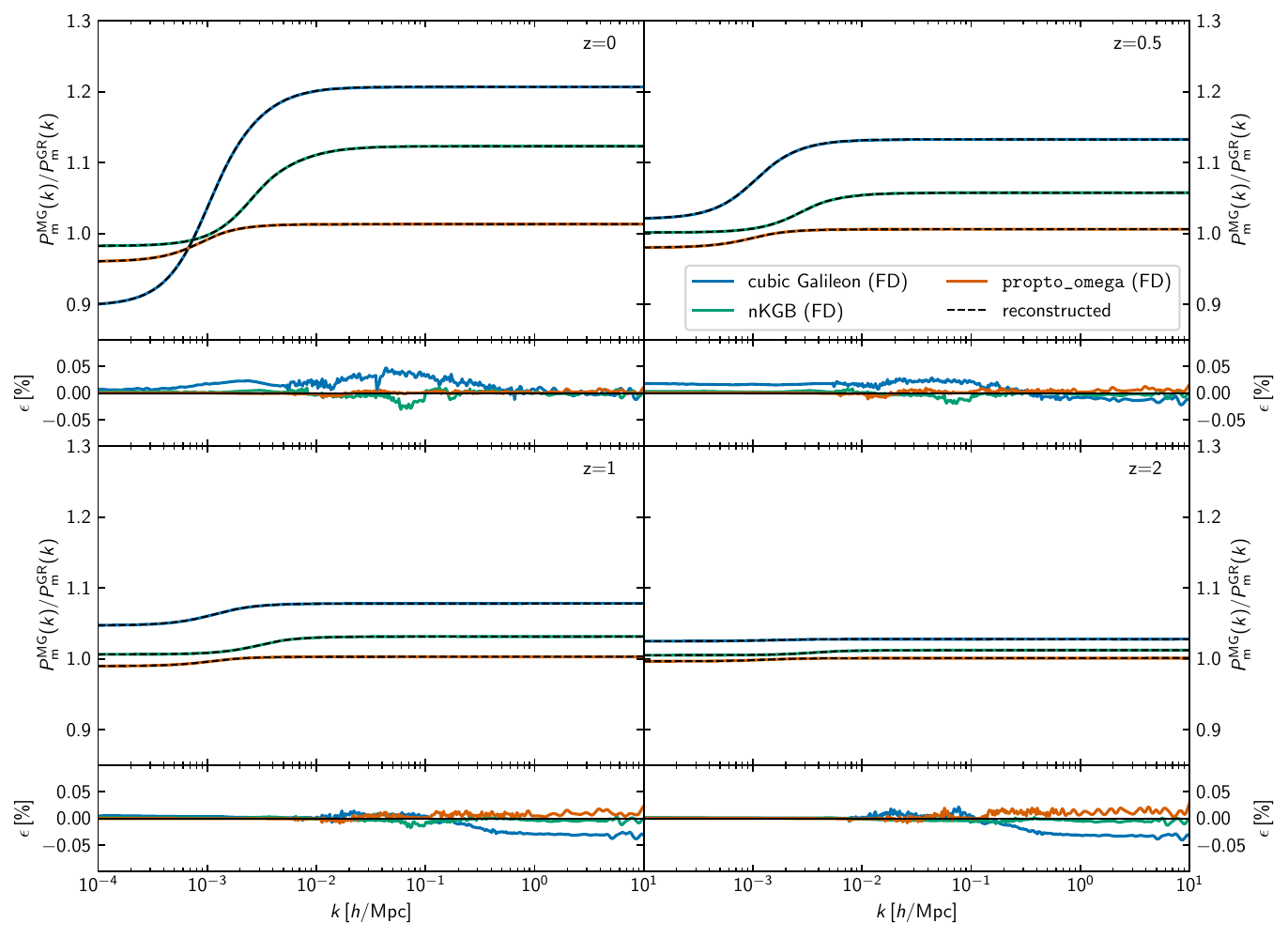}
    \caption{Evolution of the modified gravity-to-\lcdm{} power spectrum ratio for three test models with scale-independent linear growth on sub-horizon scales. The colored lines represent the output from \hiclass{} solving for the full dynamic (FD) of the scalar degree of freedom. The dashed lines are the \mochiclass{} predictions for the reconstructed models, where the stable parameters, \( \{ \DeltaMpl, \Dkin, \cs, \normrhophi \} \), and the initial conditions, \( \alphaBic \), are computed as detailed in the main text. The excellent agreement between the exact and reconstructed power spectrum ratios, shown in the lower sub-panels, demonstrates that the combination of principal components with a power-law approximation successfully captures the linear growth of structure in these models.}
    \label{fig:pk_reconstructed_scale_independent}
\end{figure*}

\begin{figure*}[ht]
    \centering
    \includegraphics[width=\textwidth]{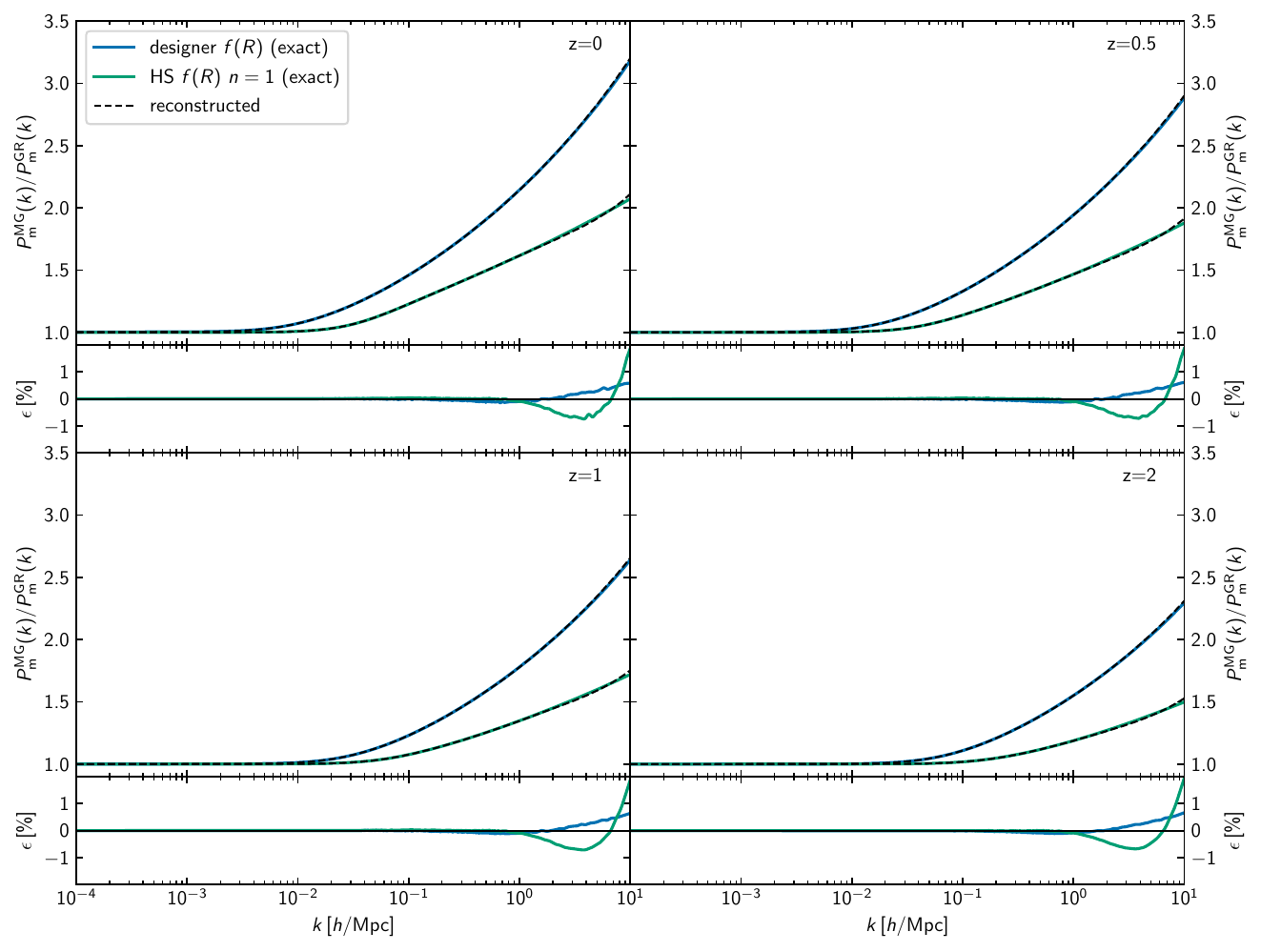}
    \caption{Evolution of the modified gravity-to-\lcdm{} power spectrum ratio for two \fr{} gravity models--designer and Hu-Sawicki (HS). The colored lines represent the output from \mochiclass{} using the exact input stable parameters, \( \{ \DeltaMpl, \Dkin, \cs, \normrhophi \} \), and initial conditions, \( \alphaBic \). The dashed lines are predictions from \mochiclass{} using the reconstructed input functions, with initial conditions optimised to compensate for reconstruction inaccuracies. The lower sub-panels show that at all examined redshifts the reconstructed power spectra remain within 0.2\% of the exact models for \( k \lesssim 1 \, \hMpc \), and within 1\% for scales up to \( k = 8 \, \hMpc \). However, reconstructing the HS model with $n=4$ (not shown) proved extremely difficult; despite sub-percent errors in the reconstructed stable functions (refer to Figs.~\ref{fig:Dkin_DeltaM_reconstruction} and~\ref{fig:rho_phi_reconstruction}), we could not determine initial conditions leading to stable perturbations.}
    \label{fig:pk_reconstructed_scale_dependent}
\end{figure*}

Figures~\ref{fig:Dkin_DeltaM_reconstruction} to \ref{fig:rho_phi_reconstruction} illustrate the reconstruction accuracy of the principal components combined with the power law and/or constant approximation for the four basis functions in all of our test models. To project the exact functions onto the principal components, we first compute the natural logarithm of their absolute values~\footnote{The stability conditions require $\Dkin$ and $\cs$ to be positive. However, $\DeltaMpl$ and $\normrhophi$ can be negative, hence the use of the absolute value.}. We then remove the power-law and/or constant trends from this quantity and finally apply the projection. For instance, to reconstruct the de-mixed kinetic term for any particular model, we do the following:
\begin{enumerate}
    \item Compute the natural logarithm of the exact $\Dkin$. 
    \item Fit a power-law over the range $[-5,-3]$ to find the best fit parameters, $\zeta_D$ and $b_D$;
    \item Subtract the power law from the exact $\ln(\Dkin)$ to obtain $\Delta_D$;
    \item Project $\Delta_D$ onto the six principal components to get $\tilde\Delta_D = \sum_i^{N_D} w_i \psi_i(x)$, where the $w_i$'s are the $N_D$ projection weights and $\psi_i$ is the $i$-th principal component.
    \item Define the approximated function using Equation~\eqref{eq:new_parametrisation} by replacing $\Delta_D$ with $\tilde\Delta_D$.
\end{enumerate}

From the lower panels of Figures~\ref{fig:Dkin_DeltaM_reconstruction}-\ref{fig:rho_phi_reconstruction} we observe that our PCA reconstruction matches the exact stable functions to within 1\% for $a \gtrsim 0.05$. In a few cases, however, the relative difference, $\epsilon$, exhibits a peak exceeding the target accuracy around $a=0.1$. Due to negligible departures from \lcdm{} at those redshifts, this is not concerning and has no measurable effect on the cosmological observables. For comparison, Figure~\ref{fig:pade_reconstruction} in Appendix~\ref{sec:pade} demonstrates that when using the same number of free parameters, alternative parametrisations of the $\Delta_J$ functions rooted in polynomial expansion underperform compared to the GP+PCA strategy adopted here.

To test how well the reconstructed stable functions describe the original MG cosmologies, we can use them as input to \mochiclass{} to predict the background expansion and linear perturbations. To assess the performance of the background reconstruction, we examine the Hubble parameter and conclude that the reconstructed expansion history resembles the exact one to better than 0.2\% for all models. However, having the full set of stable functions is not sufficient to compute the perturbations, as we also need the initial condition, $\alphaBic$, for the integration of the braiding parameter. We determine this by minimizing the mean squared error
\begin{equation}\label{eq:mse}
    {\rm MSE}[\alphaBic] = \frac{1}{N_k} \sum_k \left[ \frac{P_{\rm m}^{\rm rec.}(k | \alphaBic)}{P_{\rm m}^{\rm exact}(k)} -1 \right]^2 \, ,
\end{equation}
where $N_k$ is the total number of wavenumbers, $k$, in the interval $[10^{-4},10] \, \hMpc$, and $P_{\rm m}^{\rm rec.}(k | \alphaBic)$ and $P_{\rm m}^{\rm exact}(k)$ are the present-day matter power spectra of the reconstructed and exact model, respectively. The derived best-fit initial conditions are typically within 0.1-1\% of their exact counterparts. Due to the high sensitivity of \fr{} gravity to small inaccuracies in the reconstructed functions (mainly $\DeltaMpl$ and $\normrhophi$), we could not find a satisfactory $\alphaBic$ that produced stable perturbations for the Hu-Sawicki $n=4$ model. We do not reconstruct trivial stable functions: the speed of sound in \fr{} gravity is fixed at 1, $\DeltaMpl = 0$ for the cubic Galileon and nKGB models, and $\normrhophi = 1$ for \proptoomega{}.

Figure~\ref{fig:cmb_reconstruction} confirms that by combining GPs and PCA we can predict the secondary CMB anisotropies (i.e. lensing and ISW effect) sourced by the late-time modified growth of the test models to precision levels well within the expected uncertainties of CMB-S4 experiments~\citep[see, e.g.,][]{Ade2018}. Matter power spectra for cubic Galileon, nKGB and \proptoomega{} models are also reconstructed to $\lesssim 0.05\%$ precision across all redshifts, as shown in Figure~\ref{fig:pk_reconstructed_scale_independent}. On the other hand, even though the reconstruction accuracy of the stable functions for \fr{} gravity never exceeds 1\%, Figure~\ref{fig:pk_reconstructed_scale_dependent} highlights that the derived matter power spectra deviate from the truth by $\lesssim 0.2\%$ only up to $k \approx 1 \, \hMpc$, with the agreement degrading to 1\% at $k \approx 8 \, \hMpc$.  Despite these relatively large differences on small scales, uncertainties in the non-linear evolution~\citep{Winther2015} and baryonic feedback~\citep{Chisari2019} have a much greater impact in this regime, effectively making the reconstructed models observationally indistiguishable from the true \fr{} gravity theories.

\subsection{Exploring Horndeski's landscape}\label{sec:rnd_models}

\begin{figure*}[ht]
    \centering
    \begin{minipage}{.49\textwidth}
        \centering
        \includegraphics[width=\linewidth]{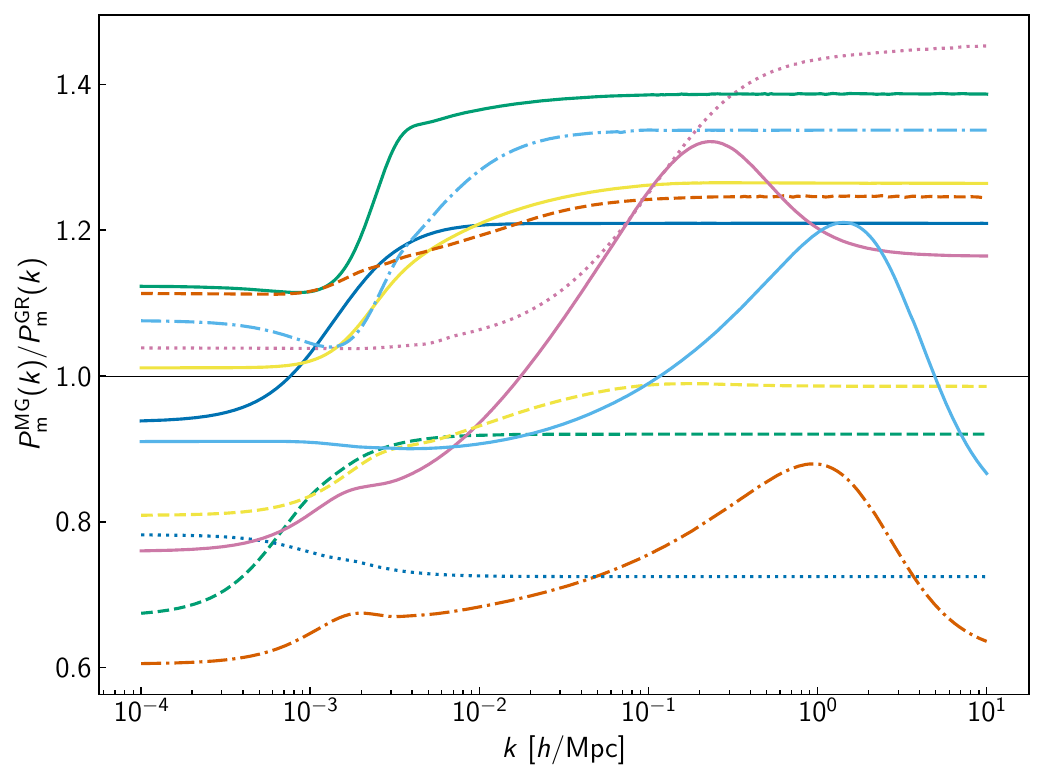}
    \end{minipage}%
    \begin{minipage}{.49\textwidth}
        \centering
        \includegraphics[width=\linewidth]{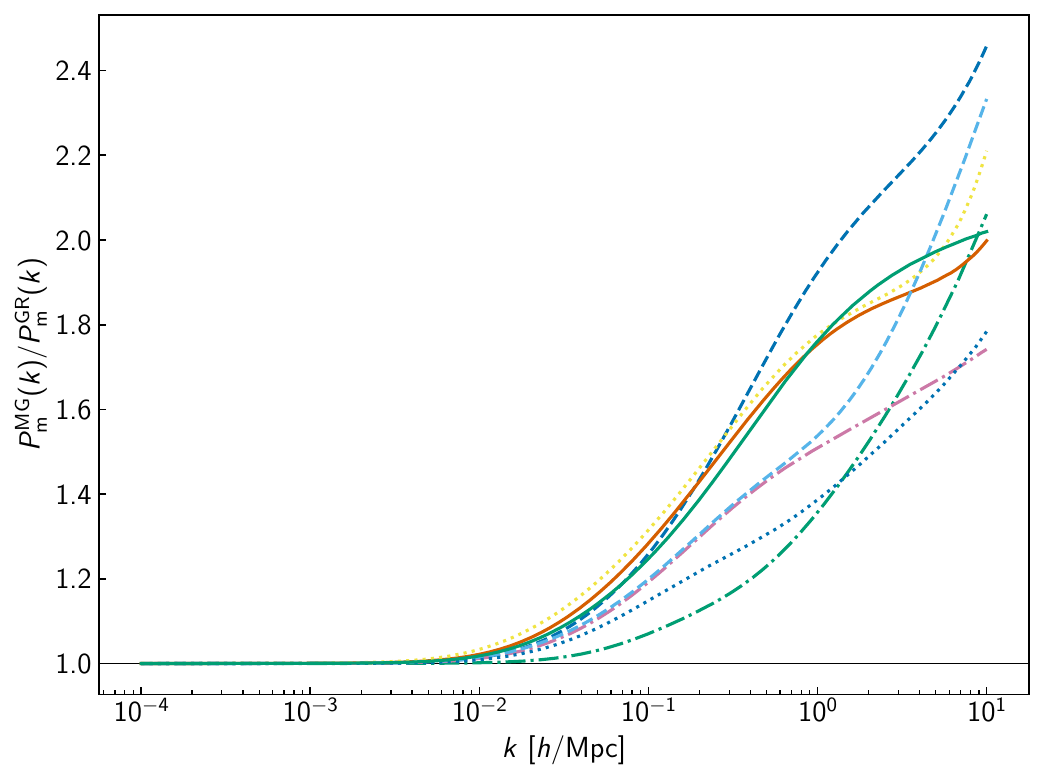}
    \end{minipage}
        \caption{Ratios of the matter power spectrum at \( z=0 \) for randomly generated modified gravity models to that of \lcdm{}. \leftcap{}: 12 stable models selected from a 31-dimensional parameter space, sampled via a Latin hypercube strategy. We use 6 principal components for each of the input functions, 2 power-law parameters to describe the early-time evolution of \( \DeltaMpl \), \(\Dkin \) and \( \normrhophi \), as well as 1 scaling parameter setting the amplitude of \( \cs \) deep in the matter-dominated era. The initial conditions, \( \alphaBic \), are determined to ensure the solution to the braiding non-linear ODE (Equation~\ref{eq:ode_braid}) closely follows the expected behaviour at early times (Equation~\ref{eq:braid_early}). The observed variety of power spectrum shapes demonstrates that the parametrisation of the stable functions proposed here can capture a richer phenomenology than the fixed-form parametrisations, like \proptoomega{}, or traditional models, such as Galileons and nKGB. \rightcap{} 8 realisations drawn from a pool of stable models obtained by sampling a 16-dimensional parameter space with a Latin hypercube strategy. In this case we only parametrise \( \DeltaMpl \) and \( \normrhophi \) using 6 principal components and 2 power-law parameters. We set \( \cs = 1 \), and \( \Dkin \) is given by \( 3\alphaM^2/2 \). Initial conditions for the braiding follow the same methodology as the left panel. In this sub-space of the full Horndeski's landscape, models with \( \alphaK \ll \alphaB^2 \) are closely related to \fr{} gravity theories. Indeed, the morphology of power spectrum ratios is similar yet more varied than the \fr{} gravity test models used in this work.}
    \label{fig:pk_rnd_models}
\end{figure*}

By combining \mochiclass{} with the newly introduced parametrisation, Equation~\eqref{eq:new_parametrisation}, we can venture beyond the phenomenological boundaries defined by the \lcdm{} extensions discussed in Section~\ref{sec:horndeski}. For this first exploratory task, we focus on two families of models:
\begin{enumerate}
    \item Scalar Horndeski (SH): here, we exploit the full functional freedom provided by Equation~\eqref{eq:new_parametrisation} and use the GP+PCA basis presented in Section~\ref{sec:gp_pca}. For simplicity, we fix the number of principal components to six for all $\Delta_J$'s, and given the small contribution of the power-law term to the speed of sound (see Figure~\ref{fig:cs2_reconstruction}) we set $A_{c_{\rm s}} = 0$. Therefore, we have a total of 31 parameters accounting for the power laws and PC weights.
    \item \fr{}-{\it like} (FR): this represents a particular sub-space of the SH cosmologies, where sampling can be extremely challenging without imposing specific relations for the stable functions. The freedom is limited to $\DeltaMpl$ and $\normrhophi$, with the remaining functions being $\cs = 1$ and $\Dkin = 3\alphaM^2/2$. Models with $\alphaK \ll \alphaB^2$ closely resemble \fr{} gravity. To describe this sub-class we use six PC weights and two power-law parameters for each stable basis function.
\end{enumerate}
In both cases we sample the parameter space using a Latin hypercube strategy with 5,000 sampling points for SH and 1,500 for FR. The parameters ranges defining the hypercube can be found in Appendix~\ref{sec:lh}.

As with the reconstruction of the test models, we need to choose a value for the initial condition, $\alphaBic$. While in the former case we could use the exact calculations as reference and minimise Equation~\eqref{eq:mse}, here we are generating new models, and $\alphaBic$ should be interpreted as an additional free parameter. However, if we want the early-time evolution of the braiding to be driven by the stable functions, we can impose an additional constraint to find suitable initial conditions. To see this, we start by linearising Equation~\eqref{eq:ode_braid} around $\alphaB = 0$. This approximation is valid for times $x \leq x_0$, with $x_0$ sufficiently deep in the matter-dominated era, when we expect $\alphaB \rightarrow 0$. By applying the integrating factor technique, we can express the solution to the linearised non-homogeneous differential equation as
\begin{equation}\label{eq:braid_linear_solution}
    \alphaB^{\rm (lin)}(x) = \left[ \mathcal{V}(x) + \alphaB(x_0) \right] e^{\mathcal{G}(x)} \, ,
\end{equation}
where
\begin{equation}\label{eq:braid_homo}
    e^{\mathcal{G}(x)} = \frac{H(x_0)}{H(x)} \frac{M_\ast^2(x)}{M_\ast^2(x_0)} e^{(x_0 - x)} 
\end{equation}
represents the time-varying part of the homogeneous solution. The function $\mathcal{V}$ contributes to the particular solution that depends on the source term, $\mathcal{S}$. It is calculated by integrating the following differential equation with the associated initial condition:
\begin{equation}\label{eq:ode_braid_particular}
    \mathcal{V}^\prime = e^{-\mathcal{G}(x)} \mathcal{S}(x) , \qquad \mathcal{V}(x_0) = 0 \, .
\end{equation}
Note that this equation contains no information about $\alphaBic$. In contrast, $\alphaB(x_0)$ does depend on the value of the braiding at $x=0$. In the regime of interest here, beside applying the approximations in Equation~\eqref{eq:deviation_limits}, we can also neglect radiation by setting $\Omega_{\rm r} = 0$. This allows us to simplify the stable functions as power laws plus constants, write $(\sfrac{H}{H_0})^2 \approx \Om e^{-3x}$, and approximate $M_\ast^2(x) \approx M_\ast^2(x_0) \approx 1$. Equation~\eqref{eq:ode_braid_particular} can then be integrated analytically for all times $x \leq x_0$, and its solution consists of two parts: a constant term, $\bar v$, that depends on $x_0$, and a sum of power laws. In the limit $x \ll x_0$, the power laws vanish, leaving $\bar v$ as the only remaining contribution. 

Now, let us assume a value $\alphaBic$ such that $\alphaB(x_0)$ is either larger or smaller than $-\bar v$. The early-time evolution of the approximate solution, Equation~\eqref{eq:braid_linear_solution}, will be controlled by the homogeneous solution, Equation~\eqref{eq:braid_homo}, which ultimately scales as $\sim  \pm \sqrt{e^{x}}$. This scaling is completely independent of the cosmological parameters and late-time dynamics defined by the stable functions. To avoid this spurious behaviour, we must adjust $\alphaBic$ such that $\alpha_{B}(x_0) = -\bar v$, which yields the following early-time solution: 
\begin{align}\label{eq:braid_early}
    \alphaB^{\rm (lin)}(x) \approx& -\sign(A_M) \frac{2\zeta_M + 3}{\zeta_M - 1/2} e^{(\zeta_M x + b_M)} \nonumber \\ 
                &+ \frac{C_{\cs}}{\zeta_D - 1/2}e^{(\zeta_D x + b_D)} \nonumber \\ 
                &+ \sign(A_\rho) \left( \frac{1 - \Om}{\Om} \right) \frac{\zeta_\rho}{\zeta_\rho + 5/2} e^{(\zeta_\rho + 3)x + b_\rho} \, .
\end{align}

As a case in point, we examine a designer \fr{} gravity cosmology with a \lcdm{} background and $B_0 \ll 1$. For this model, Equation~\eqref{eq:braid_early} simplifies significantly, as we have $\cs = 1$, $\Dkin = 3\alphaM^2/2 \ll \alphaM$, and $\normrhophi = 1$. From the definition of the running, Equation~\eqref{eq:run}, and by imposing the condition $\alphaB = -\alphaM$, we derive a quadratic equation for the slope of $\DeltaMpl$, with the positive solution being $\zeta_M = (5 + \sqrt{73})/4$. Reassuringly, this result matches the value derived by \cite{Pogosian2007} in the limit $\Omega_{\rm r} \rightarrow 0$ \citep[see also][]{Lombriser2018}.

Finally, for each set of principal component weights, constants, power-law slopes, and amplitudes, we compute the stable functions and use Equation~\eqref{eq:braid_early} as an early-time target for the integrated braiding solution to obtain $\alphaBic$. In practice, we find the zero of the function defined as the definite integral over the range $x \in [-5,-4]$ of the difference $\alphaB^{\rm (ode)} - \alphaB^{\rm (lin)}$, where $\alphaB^{\rm (ode)}$ is the solution to Equation~\eqref{eq:ode_braid}. For this step, we run only the background module of \mochiclass{} and search for $\alphaBic$ in the interval $[0, 2]$~\footnote{Although no theoretical argument prevents us from considering negative $\alphaBic$, we restrict ourselves to positive values at the cost of missing a small fraction of viable models. This choice is also supported by recent cosmological analyses, which found $\alphaBic \gtrsim 0$ with high statistical significance regardless of the employed parametrisation~\citep{Kreisch2017,Noller2018,Mancini2019,Raveri2019}.}. The root finder typically takes 6-7 iterations to reach convergence. Consequently, this process slightly degrades the performance presented in Section~\ref{sec:performance} since more time must be spent on the background calculations.

The left panel of Figure~\ref{fig:pk_rnd_models} shows the present-day matter power spectrum ratio for 12 models selected from approximately 200 stable cosmologies generated using the SH Latin hypercube. A viability rate of 4\% might seem surprising given that all 5,000 sample models pass the stability criteria, Equation~\eqref{eq:stability_conditions}. However, only a small fraction of these models feature stable perturbations not affected by exponentially growing modes, highlighting the importance of starting from the stable basis functions--using the $\alpha_i$'s to sample this parameter space would be even more challenging. Similar arguments apply to the right panel, where we find a higher stability rate of 20\%, owing to the considerably smaller parameter volume.

Returning to the left panel, the variety of shapes for the matter power spectrum ratios reflects the rich phenomenology of the SH landscape. Besides the familiar behavior displayed by models such as cubic Galileon, nKGB, and \proptoomega{}, the growth of structure can be slower than in \lcdm{}, and can also exhibit non-trivial scale dependence arising from the interplay between the braiding scale, the Compton scale, and the scalar field sound horizon~\citep[see, e.g.,][for detailed discussions]{Sawicki2012,Bellini2014}.

As expected, the power spectrum ratios shown in the right panel of Figure~\ref{fig:pk_rnd_models} generally follow the trend of our test \fr{} cosmologies (see, e.g., Figure~\ref{fig:mochi_v_xcamb_pk}). However, the particular scale dependence can depart significantly from the commonly adopted designer and Hu-Sawicki models. This behaviour can be primarily attributed to (i) a different, possibly time-dependent, gravitational coupling in the sub-Compton limit (i.e., $\mu_{\infty} \neq \mu_{\infty}^{f(R)} = 4/3$), and (ii) a more complex evolution of the Compton wavelength.

\section{Summary and outlook}\label{sec:summary}

Horndeski gravity provides a comprehensive theoretical framework to explore scalar-tensor extensions of GR, offering rich phenomenology through diverse background expansion and structure formation scenarios. To study this extensive Horndeski landscape, we can restrict ourselves to the energy scales relevant to cosmology by adopting an effective-theory approach. In this work, we have introduced a numerical tool to facilitate the study and statistical analysis of Horndeski models expressed in the effective-theory language.

At the level of the background and linear perturbations, models within the scalar Horndeski sub-class are entirely described by four functions of time. The common $\alpha$-parametrisation, proposed by \cite{Bellini2014}, neatly separates the phenomenology of the background from that of the growth of structure and is implemented in Einstein-Boltzmann solver \hiclass{}. This code also includes the quasi-static approximation scheme for improved computational efficiency and to extend applicability to models where following the full dynamics of the scalar field is challenging.

Despite these key computational advances, to reliably extend the use of \hiclass{} to a broader portion of the Horndeski landscape, we must first address the following critical points:
\begin{itemize}
    \item Simplify the implementation of specific models, which currently still requires modifications to the source code.
    \item Generalise the input functions to more flexible parametrisations, going beyond \(\{ w_0, w_a \}\) for \(\rhophi\), and \(\propto \Omega_\phi(a)\) or \(a^s\) for the \(\alpha\)-functions.
    \item Efficiently select stable models, eliminating the need to search for parameter combinations free from ghost and gradient instabilities.
    \item Improve on the accuracy of the QSA implementation in \hiclass{}.
    \item Correct the erroneous labeling of healthy models as unstable due to limited numerical precision in the calculation of \(\cs\).
\end{itemize}

The \hiclass{} extension released with this work, \mochiclass{}, effectively resolves all the points above by:
\begin{itemize}
    \item Replacing the \(\alpha\)-functions directly with the parameters used in the stability conditions, \(\{ \Dkin, \Mpl, \cs \}\). To ensure the absence of ghosts and gradient instabilities, users need only to provide positive input functions.
    \item Accepting general functions of time as input, with the constraints \(\csnum \rightarrow 0\), \(\Mpl \rightarrow 1\), and \(\rhophi \ll \rhom\) deep in the matter-dominated era. In other words, the model must resemble the \lcdm{} cosmology at early times.
    \item Including mathematical (or classical) stability conditions to catch exponentially growing modes, as done in \eftcamb{}.
    \item Embedding a QSA implementation based on modified Poisson-{\it like} equations for the metric potentials.
\end{itemize}

We validated \mochiclass{}'s accuracy and performance by comparing its output power spectra and computational efficiency to those of the publicly available Einstein-Boltzmann solvers \hiclass{}, \eftcamb{}, and \mgcamb{}. For all tested cases, we found that the predicted changes to the CMB spectra due to new physics matched the output from the other solvers within 0.05\%, while for the matter power spectrum, the agreement reached 0.01\%. Thanks to its alternative QSA implementation and to the manifestly stable EFT basis, \mochiclass{} generally improves on the accuracy and numerical stability attained by \hiclass{}, with \( f(R) \) gravity being an extreme case that \hiclass{} cannot even solve.

In addition, we proposed a new parametrisation for the stable functions and the background energy-density of the scalar field. This approach uses a simple early-time power-law (or constant) evolution and models late-time smooth deviations with Gaussian Processes, combined with principal component analysis for dimensionality reduction. We showed that this parametrisation can accurately reconstruct our test cosmologies, both in terms of input functions and output power spectra. As an initial application, we used it to generate Horndeski models with either all input functions free or by fixing relations and values to closely resemble \fr{} gravity. The linear growth of structure in these models can vary significantly, deviating from the well-known behavior of our test models. This diversity emphasises the necessity of exploring beyond the commonly employed parametrisations~\citep[see also][for a discussion on biased constraints and information loss]{Raveri2021}.  

The high versatility of \class{}'s Python wrapper, \texttt{classy}, ensures \mochiclass{} seamless integration into the pipeline of cosmological analyses using data probing both the background evolution and the linear regime of structure formation. This will provide an alternative avenue of investigation to the reconstruction strategies employed in, e.g., \cite{Raveri2019,Zucca2019b} and \cite{Pogosian2021,Raveri2021}, and will extend the analysis of studies based on fixed-form parametrisations of the $\alpha$- or EFT-functions~\citep[e.g.,][]{Salvatelli2016,Noller2018,Aghanim2018,Seraille2024}. In a future release, we will also include a special case of the stable parametrisation enforcing the relation $\alphaB = b\alphaM$, which can be more effective for analyses restricted to models in the generalised Brans-Dicke sub-class~\citep{Bergmann1968}, where $\alphaB = -\alphaM$, or to investigate no-slip gravity modifications~\citep{Linder2018}, defined by $\alphaB = -2\alphaM$. 

The larger volume of Horndeski theory space covered by the new parametrisation will likely require scalable sampling methods capable of efficiently handling the high dimensionality of the problem~\citep{Mootoovaloo2024}. One such method is Hamiltonian Monte Carlo~\citep[e.g.,][]{Hajian2006}, a gradient-based sampling technique that leverages the derivatives of the likelihood function to significantly enhance the acceptance rate of samples in high-dimensional spaces. To implement this, we could rewrite \mochiclass{} using the \texttt{JAX} framework, taking advantage of its automatic differentiation capabilities, as demonstrated by~\cite{Hahn2023}. Alternatively, a simpler approach might involve using \mochiclass{} to generate training data for neural network emulators targeted to specific summary statistics~\citep{Mancini2021,Bonici2023}. These emulators can then be integrated into cosmology libraries that feature automatic differentiation~\citep{Piras2023,Campagne2023}. 

Finally, for general Horndeski models, \mochiclass{}'s calculations represent a fundamental first step towards accessing the cosmological information contained on small scales~\citep{Heymans2018}. For instance, in the halo model reaction framework~\citep{Cataneo2018,Giblin2019}, the linear matter power spectrum and background evolution are key ingredients for obtaining the mean halo statistics and non-linear matter power spectrum of non-standard cosmologies. This method, implemented in the \texttt{ReACT} code~\citep{Bose2020}, can be applied to Lagrangian-based theories of gravity~\citep{Cataneo2018,Bose2021,Atayde2024,Bose2024}, to phenomenological modifications to GR~\citep{Srinivasan2021,Srinivasan2023}, to an interacting dark sector~\citep{Carrilho2021}, and to a broader class of parametrised \lcdm{} extensions within the Horndeski realm~\citep{Bose2022}.

\acknowledgements
\section*{Acknowledgements}

We acknowledge the other \hiclass{} developers, Miguel Zumalac\'arregui and Ignacy Sawicki, for sharing a private version of the code.
We are grateful to Noemi Frusciante and Alessandra Silvestri for their help with \texttt{EFTCAMB}. We would also like to thank Lucas Lombriser for insightful discussions on the inherently stable parametrisation, and Lucas Porth for brainstorming strategies to optimise initial conditions. Finally, a special thank to Alexander Eggemeier for suggesting a memorable name for the code released with this work. We acknowledge the use of the software packages \texttt{numpy}, \texttt{scipy}, \texttt{matplotlib}, and \texttt{scikit-learn}. Additionally, \texttt{Mathematica} was utilized for comparison purposes and to compute the stable basis functions of some test cosmologies. MC acknowledges the sponsorship provided by the Alexander von Humboldt Foundation in the form of a Humboldt Research Fellowship for part of this work. He also acknowledges support from the Max Planck Society and the Alexander von Humboldt Foundation in the framework of the Max Planck-Humboldt Research Award endowed by the Federal Ministry of Education and Research. EB~ has received funding from the European Union’s Horizon 2020 research and innovation program under the Marie Skłodowska-Curie grant agreement No 754496.

\clearpage

\appendix

\section{Precision parameters used in \class{} and \camb{}}\label{sec:precision}

The level of agreement between \class{} and \camb{} strongly depends on the precision settings adopted in the two codes~\citep{Lesgourgues2011a}. To minimize the impact of numerical errors on the power spectra produced by different implementation strategies, code comparison analyses often employ high-precision settings~\citep{Bellini2017}. Following this philosophy, we use high-precision parameter values for both \mochiclass{} and \hiclass{},   
\begin{verbatim}
    start_small_k_at_tau_c_over_tau_h = 1e-4
    start_large_k_at_tau_h_over_tau_k = 1e-4
    perturbations_sampling_stepsize = 0.01
    l_logstep = 1.026
    l_linstep = 25
    k_max_tau0_over_l_max = 8.
    k_per_decade_for_pk = 200
    l_max_scalars = 5000
    P_k_max_h/Mpc = 12.
    delta_l_max = 1000
    l_switch_limber = 50
\end{verbatim}
Similarly for \mgcamb{} and \eftcamb{} we use
\begin{verbatim}
    get_transfer = T 
    transfer_high_precision = T 
    high_accuracy_default = T 
    k_eta_max_scalar = 80000 
    do_late_rad_truncation = F 
    accuracy_boost = 2 
    l_accuracy_boost = 2 
    l_sample_boost = 2 
    l_max_scalar = 10000 
    accurate_polarization = T 
    accurate_reionization = T 
    lensing_method =1 
    use_spline_template = T
\end{verbatim}
In addition, for \eftcamb{} we set
\begin{verbatim}
    EFTCAMB_turn_on_time = 1e-10
    EFTtoGR = 1e-10
\end{verbatim}
and for \mgcamb{} we use the default
\begin{verbatim}
    GRtrans = 1e-3
\end{verbatim}

\section{Reconstruction with Pad\'e approximants}\label{sec:pade}

\begin{figure*}[ht]
    \centering
    \begin{minipage}{.49\textwidth}
        \centering
        \includegraphics[width=\linewidth]{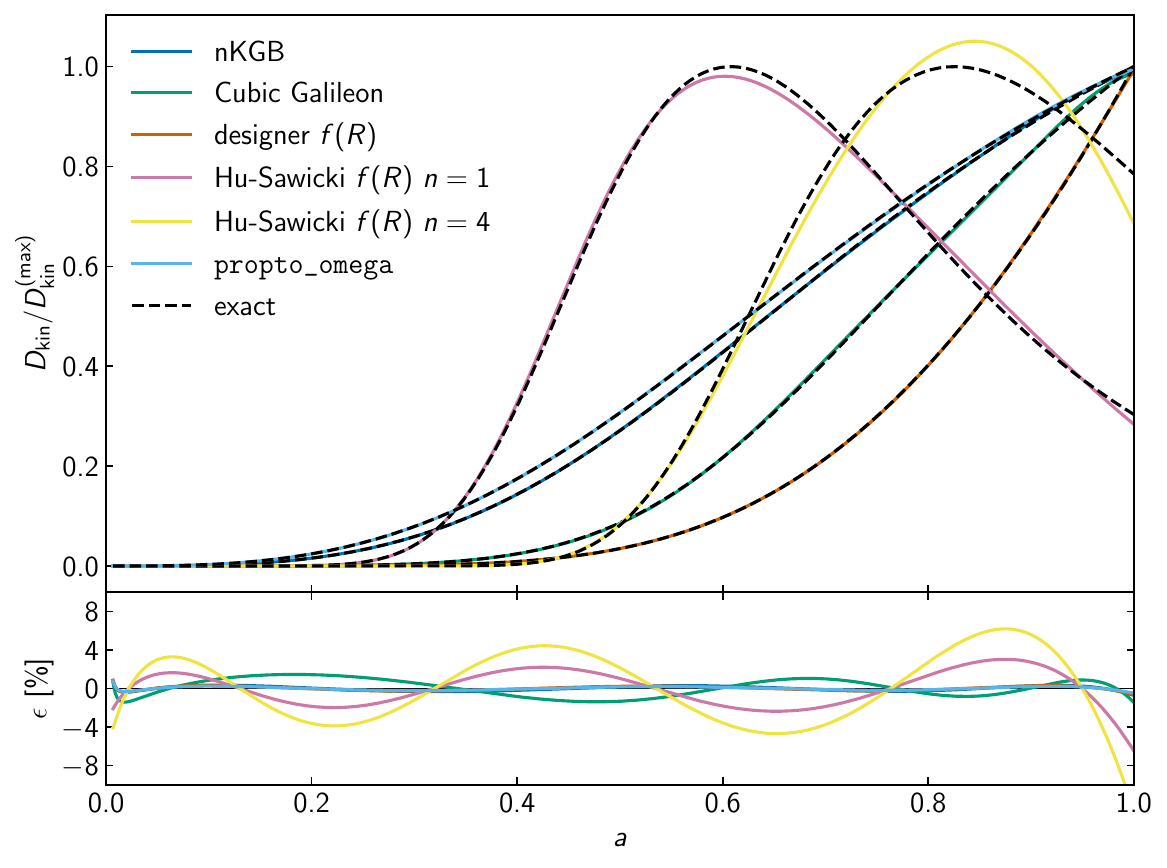}
    \end{minipage}%
    \begin{minipage}{.49\textwidth}
        \centering
        \includegraphics[width=\linewidth]{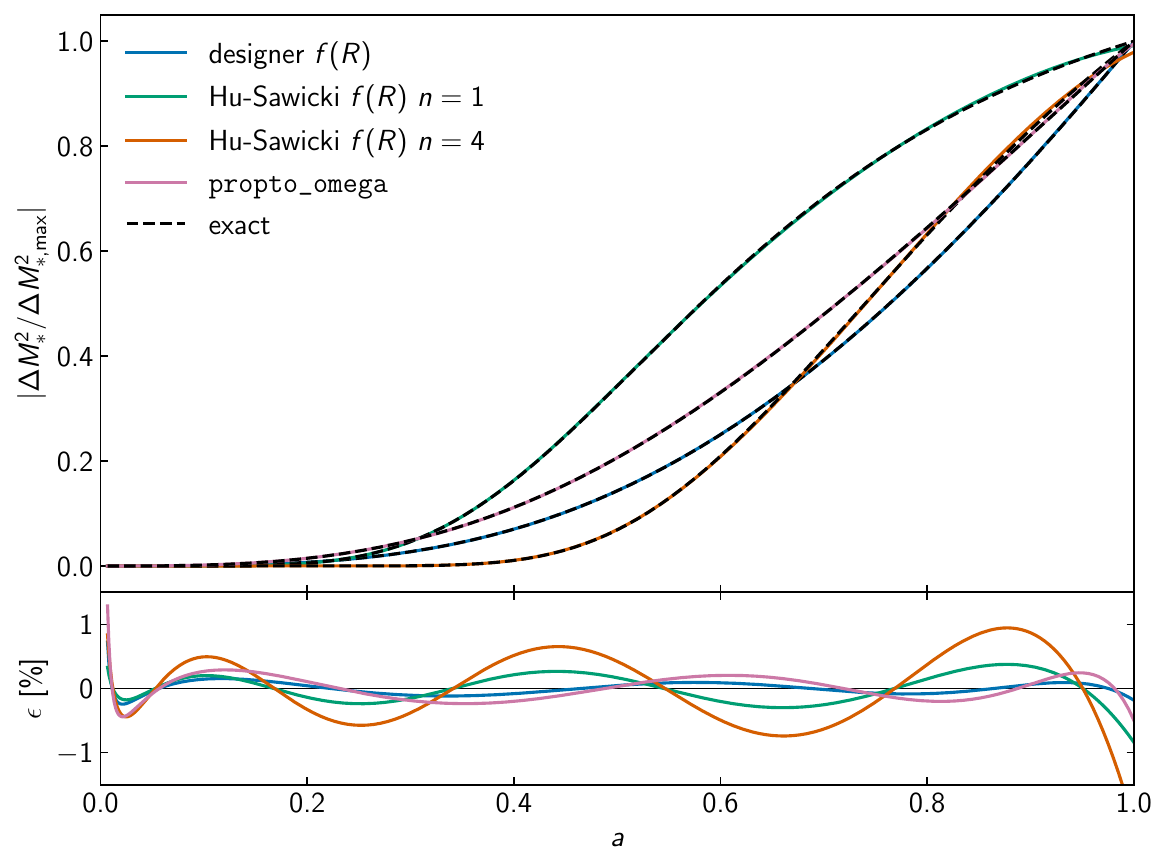}
    \end{minipage}
    \caption{Reconstruction of two stable basis functions using Pad\'e approximants (Equation~\ref{eq:pade}). The left panel illustrates the normalized de-mixed kinetic term (\(D_{\text{kin}}/D_{\text{kin}}^{\text{(max)}}\)), while the right panel displays the normalized absolute deviations from the standard cosmological strength of gravity (\( |\Delta M^2/\Delta M^2_{\ast,\text{max}}| \)). The coloured lines represent functions reconstructed by fitting Padé approximants to their exact counterparts (dashed lines), once the early-time power-law trend has been subtracted. We maintain an identical number of free parameters as in the PCA-based approach to ensure a fair comparison. The lower sub-panels show the reconstruction accuracy of the fittings, indicating that the Pad\'e approximants result in larger deviations from the exact models compared to the PCA-based reconstructions.}
    \label{fig:pade_reconstruction}
\end{figure*}

A Pad\'e approximant of order \([N/M]\) for a function \(f(x)\) is given by the ratio of two polynomials of degrees \(N\) and \(M\),
\begin{equation}\label{eq:pade}
    Q(x)=\frac{\sum_{n=0}^{N}\alpha_{n}\left( x - x_0 \right)^{n}}{1+\sum_{m=1}^{M}\beta_{m}\left( x - x_0 \right)^{m}}\;,
\end{equation}
where the coefficients \(\alpha_n\) and \(\beta_m\) are chosen such that the Taylor series expansion of \(Q(x)\) around \(x = x_0\) matches the Taylor expansion of \(f(x)\) up to order $N+M$. 

Here, we quantify the reconstruction accuracy of the Padé approximants for the deviations, \(\Delta_J\), used in the new parametrisation of the stable functions described in Section~\ref{sec:new_models}. We perform the expansions around \(x_0 = 0\), but have verified that our conclusions are not sensitive to this particular choice. Since we lack analytical expressions for the \(\Delta_J\) of our test models, we cannot derive the Padé coefficients by matching the Taylor series of \(\Delta_J\) order by order. Instead, we interpret Equation~\eqref{eq:pade} as a fitting function and determine the best-fit values for \(\alpha_n\) and \(\beta_m\) by minimising the reconstruction error. To ensure a fair comparison with the GP + PCA strategy discussed in Section~\ref{sec:gp_pca}, we use the same number of free parameters by setting \(N = 2\) and \(M = 3\), except for \(\Delta_M\), where \(N = 3\). 

Figure~\ref{fig:pade_reconstruction} illustrates the performance of the Padé approximants for two representative cases: \(\Dkin\) and \(\DeltaMpl\). Reconstruction errors can reach approximately 8\% for the de-mixed kinetic term and around 1\% for the Planck mass shift. In comparison, Figure~\ref{fig:Dkin_DeltaM_reconstruction} demonstrates that our GP + PCA approach can match the exact quantities to better than 1\%, which motivates the use of this reconstruction technique.

\section{Latin hypercube boundaries}\label{sec:lh}

Table~\ref{tab:bounds_sh} and~\ref{tab:bounds_fr} list the parameter ranges defining the Latin hypercubes used to generate random models belonging to the Scalar Horndeski and \fr{}-{\it like} classes, respectively (see Section~\ref{sec:rnd_models}).

{
\renewcommand{\arraystretch}{1.5}

\begin{table}[ht]
    \centering
    \begin{tabular}{*{10}{c}} 
        \toprule
                & $\zeta_J$ & $b_J$ & $C_J$ & $w_1$ & $w_2$ & $w_3$ & $w_4$ & $w_5$ & $w_6$ \\
        \midrule
        \midrule
        $\DeltaMpl$ & $[2, 15]$ & $[-6,3]$ & -- & $[0,250]$ & $[0,150]$ & $[0,50]$ & $[0,10]$ & $[0,5]$ & $[-1,0]$ \\
        $\Dkin$ & $[2, 20]$ & $[-5,5]$ & -- & $[-35,0]$ & $[-50,0]$ & $[0,5]$ & $[0,5]$ & $[-2,2]$ & $[-1,1]$ \\
        $\cs$ & undefined & $-\infty$ & $[0,2]$ & $[0,30]$ & $[-30,0]$ & $[-15,0]$ & $[0,3]$ & $[-1,1]$ & $[-1,1]$ \\
        $\normrhophi$ & $[-1, 3]$ & $[-2,2]$ & derived & $[-10,10]$ & $[-8,8]$ & $[-4,4]$ & $[-1,1]$ & $[-1,1]$ & $[-1,1]$ \\
        \bottomrule
    \end{tabular}
    \caption{Parameter ranges used to generate the 5,000 Scalar Horndeski sample models following the parametrisation in Equation~\eqref{eq:new_parametrisation} combined with the PCA basis functions described in Section \ref{sec:gp_pca}. We fix $\sign(A_{M}) = -1$ and $\sign(A_{\rho}) = +1$.}
    \label{tab:bounds_sh}
\end{table}

}

{
\renewcommand{\arraystretch}{1.5}

\begin{table}[ht]
    \centering
    \begin{tabular}{*{10}{c}} 
        \toprule
                & $\zeta_J$ & $b_J$ & $C_J$ & $w_1$ & $w_2$ & $w_3$ & $w_4$ & $w_5$ & $w_6$ \\
        \midrule
        \midrule
        $\DeltaMpl$ & $[4, 15]$ & $[-6, -2]$ & -- & $[0, 250]$ & $[0, 50]$ & $[0, 10]$ & $[0, 5]$ & $[0, 5]$ & $[-1,0]$ \\
        $\normrhophi$ & 0 & $[-0.1,0.1]$ & derived & $[-1,1] \cdot 10^{-3}$ & $[-1,1] \cdot 10^{-3}$ & $[-5, 5] \cdot 10^{-4}$ & $[-1, 1] \cdot 10^{-4}$ & $[-1, 1] \cdot 10^{-4}$ & $[-1, 1] \cdot 10^{-4}$ \\
        \bottomrule
    \end{tabular}
    \caption{Parameter ranges used to generate the 1,500 \fr{}-{\it like} sample models following the parametrisation in Equation~\eqref{eq:new_parametrisation} combined with the PCA basis functions described in Section \ref{sec:gp_pca}. We fix $\sign(A_{M}) = -1$ and $\sign(A_{\rho}) = +1$.}
    \label{tab:bounds_fr}
\end{table}

}

\section{Useful equations in \mochiclass{}}\label{sec:useful_eqns}

\subsection{Continuity equation}

When users select \texttt{expansion\_model = wext}, the scalar field background energy-density, \(\rhophi\), is integrated using the following continuity equation with the corresponding initial condition:
\begin{equation}\label{eq:continuity_phi}
    \dot\rho_\phi = -3\rhophi(1 + w_\phi) \, , \qquad \rho_{\phi}(0) = \Omega_\phi H_0^2 \, .  
\end{equation}
Here we remind the reader that overdots denote derivatives with respect to \(\ln a\), and primes (used below) represent derivatives with respect to conformal time. The choice between \(\ln a\) or \(\tau\) as the time variable is dictated by the original \class{} structure, with the former used in the \texttt{background} module and the latter in the \texttt{perturbations} module.

\subsection{Scalar field fluctuations}

The coefficients and the source term in Equation~\eqref{eq:vx_pert_schematic} can be found in Appendix A.2 of ~\cite{Zumalacarregui2016}. For convenience, we provide them here:
\begin{subequations}
    \begin{align}
        &\mathcal{A}(\tau) = \Dkin (2-\alphaB) \, , \\
        &\mathcal{B}(\tau) = 8 aH \lambda_7 \, , \\
        &\mathcal{C}(\tau) = -8 a^2H^2 \lambda_8 \, , \\
        &\mathcal{D}(\tau) = 2\csnum \, , \\
        &\mathcal{E}(\tau,k) = \frac{2\csnum}{aH}k^2\eta + \frac{3a}{2H \Mpl} \left[ 2\lambda_2 \delta\rhom - 3\alphaB(2-\alphaB)\delta\presm \right] \, ,
    \end{align}
\end{subequations}
where 
\begin{subequations}
    \begin{align}
        &\lambda_2 = -\frac{3(\rhom + \presm)}{H^2\Mpl} - (2-\alphaB)\frac{H^\prime}{aH^2} + \frac{\alphaB^\prime}{aH} \, , \\
        &\lambda_7 = \frac{\Dkin}{8}(2-\alphaB) \left[ 4 + \alphaM + \frac{2H^\prime}{aH^2} + \frac{\Dkin^\prime}{aH\Dkin} \right] + \frac{\Dkin}{8}\lambda_2 \, , \\
        &\lambda_8 = -\frac{\lambda_2}{8} (\Dkin - 3\lambda_2 + \frac{3\alphaB^\prime}{aH}) + \frac{1}{8} (2-\alphaB) \left[ (3\lambda_2 - \Dkin)\frac{H^\prime}{aH^2} - \frac{9\alphaB\presm^\prime}{2aH^3\Mpl} \right] \nonumber \\
        & \quad\quad \, - \frac{\Dkin}{8}(2-\alphaB) \left[ 4 + \alphaM + \frac{2H^\prime}{aH^2} + \frac{\Dkin^\prime}{aH\Dkin} \right] \, .
    \end{align}
\end{subequations}

\subsection{Quasi-static approximation}

In the effective-field formulation of the quasi-static approximation, the gravitational coupling for the non-relativistic matter and the gravitational slip can be expressed as in Equation~\eqref{eq:efe_qsa}. The small-scale limits can be expressed in terms of the $\alpha$'s as 
\begin{equation}
    \mu_\infty = \frac{2\csnum + (\alphaB + 2\alphaM)^2}{2\csnum M_\ast^2} \, , \qquad  
    \mu_{Z, \infty} = \frac{2\csnum + \alphaB(\alphaB + 2\alphaM)}{2\csnum M_\ast^2} \, ,
\end{equation}
while
\begin{equation}
    \mu_{\rm p} = \frac{9 \left\{ aH \left[ 2\csnum + (\alphaB - 2)\alphaB + 4(\alphaB-1)\alphaM \right] 
                (\rhom + \presm + \rhophi + \presphi) + 2\alphaB(\presm^\prime + \presphi^\prime) \right\}}{4aH^3} \, .
\end{equation}  

Here, we also present a typo-corrected equation for the metric perturbations in synchronous gauge, $\eta$, used in our \mgcamb{} patch to \hiclass{}:
\begin{align}
    \eta^\prime &= \frac{1}{2}\frac{a^{2}}{9\rhom a^{2}\mu\gamma(1+w_{\rm m})/2+k^{2}}
            \left\{3\rhom(1+w_{\rm m})\mu\gamma\theta\left[1+3 \left( \frac{{\mathcal{H}}^{2}-{\mathcal H}^\prime}{k^{2}} \right) \right]
            + 3\rhom\Delta_{\rm m}\left[{\mathcal H}\mu(\gamma-1)-\mu^\prime\gamma-\mu^\prime{\gamma}\right] \right. \nonumber \\ 
            &+ 9\mu(1-\gamma)\rhom(1+w_{\rm m})\sigma_{\rm m}^\prime
            + k^{2}\alpha\left[3\rhom\mu\gamma(1+w_{\rm m})-2\left(\frac{\mathcal{H}^{2}-\mathcal{H}^\prime}{a^{2}}\right)\right]
            + 9{\mathcal H}\mu(\gamma-1)(1+w_{\rm m})\rhom\sigma_{\rm m}(3w_{\rm m}+2) \nonumber \\
            &\left. - 9(1+w_{\rm m})\rhom\sigma_{\rm m}\left[\mu^\prime(\gamma-1)+\gamma^\prime\mu+\mu(\gamma-1)\frac{w_{\rm m}^\prime}{1+w_{\rm m}}\right] \right\} \, ,
\end{align}
where $\mathcal{H} = aH$ is the conformal Hubble parameter. The corrections compared to Equation (26) in \cite{Zucca2019a} include removing an extra power of two from $\mathcal{H^\prime}$ in the first line, and changing the minus sign to a plus sign in front of the $\gamma^\prime \mu$ term in the last line.


\clearpage

\bibliographystyle{aasjournal.bst}
\bibliography{references.bib}

\end{document}